\documentclass[a4paper,11pt]{article}

\pdfoutput=1

\usepackage{jheppub}
\usepackage{amsmath}
\usepackage{xspace}
\usepackage{hyperref}
\usepackage{enumitem}
\usepackage{soul}

\newcommand{\eq}[1]{eq.~\eqref{eq:#1}}
\newcommand{\eqs}[2]{eqs.~\eqref{eq:#1} and \eqref{eq:#2}}
\renewcommand{\sec}[1]{sec.~\ref{sec:#1}}
\newcommand{\secs}[2]{secs.~\ref{sec:#1} and \ref{sec:#2}}

\newcommand{\fig}[1]{fig.~\ref{fig:#1}}
\newcommand{\figs}[2]{figs.~\ref{fig:#1} and \ref{fig:#2}}
\newcommand{\nnu}{\nonumber\\}
\newcommand{\nn}{\nonumber}
\newcommand{\bef}{\begin{figure}[t]\centering}
\newcommand{\eef}{\end{figure}}
\def\bea#1\eea{\begin{align}#1\end{align}}
\def \be  {\begin{equation}}
\def \ee  {\end{equation}}
\def \ba  {\begin{eqnarray}}
\def \ea  {\end{eqnarray}}

\newcommand{\f}{\frac}
\newcommand{\ord}[1]{\mathcal{O}(#1)}

\newcommand{\df}{\mathrm{d}}
\newcommand{\img}{\mathrm{i}}

\newcommand{\sdt}{\!\cdot\!}

\newcommand{\al}{\alpha}
\newcommand{\bt}{\beta}
\newcommand{\ga}{\gamma}

\newcommand{\de}{\delta}
\newcommand{\eps}{\epsilon}

\newcommand{\w}{\omega}
\newcommand{\si}{\sigma}

\newcommand{\cG}{{\mathcal G}}
\newcommand{\GG}{{\mathcal G}}

\newcommand{\cJ}{{\mathcal J}}

\newcommand{\bn}{\bar{n}}

\newcommand{\bnslash}{\bar{n}\!\!\!\slash}

\newcommand\as{\alpha_s}
\newcommand{\lqcd}{\Lambda_\mathrm{QCD}}

\newcommand{\one}{{(1)}}

\newcommand{\kt}{anti-k$_T$\xspace}
\newcommand{\siJF}{siJF\xspace}

\allowdisplaybreaks[2]

\title{The Energy Distribution of Subjets and the Jet Shape}

\author[a,b,c]{Zhong-Bo Kang,}
\author[c,d]{Felix Ringer}
\author[e,f]{and Wouter J. Waalewijn}

\affiliation[a]{Department of Physics and Astronomy, University of California, Los Angeles, CA 90095, USA}
\affiliation[b]{Mani L. Bhaumik Institute for Theoretical Physics, University of California, Los Angeles, CA 90095, USA}
\affiliation[c]{Theoretical Division, Los Alamos National Laboratory, Los Alamos, NM 87545, USA}
\affiliation[d]{Nuclear Science Division, Lawrence Berkeley National Laboratory, Berkeley, CA 94720, USA}
\affiliation[e]{Institute for Theoretical Physics Amsterdam and Delta Institute for Theoretical Physics, University of Amsterdam, Science Park 904, 1098 XH Amsterdam, The Netherlands}
\affiliation[f]{Nikhef, Theory Group, Science Park 105, 1098 XG, Amsterdam, The Netherlands}
                                         
\emailAdd{zkang@physics.ucla.edu}
\emailAdd{fmringer@lbl.gov}
\emailAdd{w.j.waalewijn@uva.nl}

\abstract{We present a framework that describes the energy distribution of subjets of radius $r$ within a jet of radius $R$. We consider both an inclusive sample of subjets as well as subjets centered around a predetermined axis, from which the jet shape can be obtained. For $r \ll R$ we factorize the physics at angular scales $r$ and $R$ to resum the logarithms of $r/R$. For central subjets, we consider both the standard jet axis and the winner-take-all axis, which involve double and single logarithms of $r/R$, respectively. All relevant one-loop matching coefficients are given, and an inconsistency in some previous results for cone jets is resolved. Our results for the standard jet shape differ from previous calculations at next-to-leading logarithmic order, because we account for the recoil of the standard jet axis due to soft radiation. Numerical results are presented for an inclusive subjet sample for $pp \to {\rm jet}+X$ at next-to-leading order plus leading logarithmic order.}

\preprint{NIKHEF 2017-003}

\begin{document}
\maketitle

\section{Introduction}
\label{sec:intro}

In this paper, we study the energy distribution of subjets with radius $r$ inside a jet of radius $R$, as illustrated in fig.~\ref{fig:jet-jet}. We will consider jets defined through the \kt algorithm~\cite{Cacciari:2008gp} or an infrared and collinear-safe cone algorithm. Subjets are obtained by reclustering the particles inside the reconstructed jet with radius parameter $r<R$. In addition, we consider the subjet of radius $r$ centered around the standard jet axis or the winner-take-all axis (WTA)~\cite{Bertolini:2013iqa}. We mostly focus on an inclusive jet sample $pp\to {\rm jet}+X$, but briefly discuss how our framework can be extended when a veto on additional jets is imposed.

\begin{figure}[t]\centering
\includegraphics[width=0.9in]{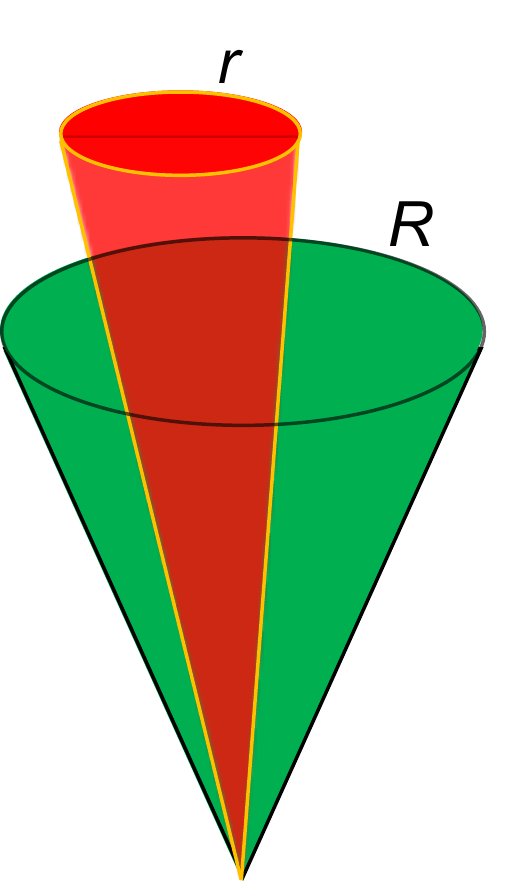}
\caption{Illustration of a subjet with radius $r$ (red) inside a jet with a radius $R$ (green). We focus on describing the energy fraction $z_r$ of the jet that is carried by the subjet.}
\label{fig:jet-jet}
\eef

Specifically, we will develop the theoretical framework to calculate
\bea \label{eq:obs}
F(z_r, r; \eta, p_T, R) = \frac{\df\sigma}{\df\eta\, \df p_T\,\df z_r} \bigg/ \frac{\df\sigma}{\df\eta\, \df p_T}
\,,\eea
where $z_r$ is fraction of the jet energy contained in the subjet of radius $r$, and $\eta$ and $p_T$ are the rapidity and transverse momentum of the jet with radius $R$. First proposed in ref.~\cite{Dai:2016hzf}, the subjet distribution simultaneously provides information about the longitudinal and transverse energy distribution inside jets, through $z_r$ and $r$. The subjet observables considered in this work are connected to both the standard jet shape~\cite{Ellis:1992qq,Seymour:1997kj,Li:2011hy,Chien:2014nsa} and the jet fragmentation function~\cite{Procura:2009vm,Jain:2011xz,Arleo:2013tya,Kaufmann:2015hma}. On the one hand, the jet shape is the average value of $z_r$ as function of $r$, for subjets centered on the jet axis. On the other hand, the jet fragmentation function describes the longitudinal momentum (or energy) fraction of hadrons in the jet. This is the $r \to 0$ limit of the inclusive subjet energy fraction $z_r$, where the collinear singularity is now cut off by hadronization instead of $r$. The jet shape~\cite{Aad:2011kq,Chatrchyan:2012mec,Chatrchyan:2013kwa,ALICE:2014dla} and the jet fragmentation function~\cite{Chatrchyan:2012gw,Chatrchyan:2014ava,ATLAS:2015mla} have been measured by the LHC experimental collaborations in both proton-proton and heavy-ion collisions. We expect that similar measurements are feasible for the observables discussed in this work.

There are several ways in which subjet distributions are valuable to present day collider phenomenology: Firstly, the various distributions of subjets discussed in this work provide a powerful test of our understanding of perturbative QCD at very high energies. Studying the energy distribution of subjets probes both the longitudinal and the transverse momentum distribution within jets at a more differential level, and may extend our current understanding of the underlying QCD dynamics. Secondly, the distribution of subjets can be used for discriminating QCD jets from boosted heavy objects, such as $W$ bosons or top quarks. Their hadronic decays would produce two or three jets, which become the subjets of one fat jet due to their boost. This plays an important role in many searches for Beyond the Standard Model (BSM) physics~\cite{Abdesselam:2010pt,Altheimer:2012mn,Altheimer:2013yza,Adams:2015hiv}. Many of the taggers used for identifying such a two or three prong decay are quite sensitive to soft radiation~\cite{Butterworth:2008iy,Plehn:2009rk,Thaler:2010tr,Larkoski:2013eya,Larkoski:2014wba,Larkoski:2014gra}. For several of the subjet observables we consider, this soft sensitivity is power suppressed and collinear factorization is sufficient. In addition to being theoretically more robust, a reduced sensitivity to soft radiation is also advantageous experimentally due to the messy LHC environment. Our work on the inclusive subjet distribution and the distribution of subjets centered about a specified axis provides a first step on the direction of taggers that are less sensitive to soft radiation. An example of a more direct connection to BSM searches is given in ref.~\cite{Rentala:2013uaa}, where the authors proposed to use jet shapes (``jet energy profiles'') to search for new physics at the LHC. Another application in the context of jet substructure is the discrimination of quark and gluon jets using e.g.~fractal observables defined on subjets rather than hadrons~\cite{Elder:2017bkd}. Thirdly, the subjet distribution is particularly suited for measuring the modification of jets in heavy-ion collisions, see e.g.~\cite{Vitev:2008rz,Chien:2015hda,Chang:2016gjp}. Jets that traverse the quark-gluon plasma get modified both in the longitudinal and transverse momentum direction, as can be collectively seen from the modification of longitudinal jet fragmentation function~\cite{Chatrchyan:2014ava} and transverse jet energy profile~\cite{Chatrchyan:2013kwa}. By identifying subjets inside a reconstructed jet, the correlations between these effects can be studied in a single measurement. 
An advantage of using subjets over hadrons is that the subjet distribution does not require the additional non-perturbative input of fragmentation functions. 

Our setup relies on collinear factorization: first we exploit that the radius $R$ is small, to factorize the dynamics of the jet from the rest of the cross section.\footnote{In practice this still works for rather large values of $R$. E.g.~in refs.~\cite{Jager:2004jh,Mukherjee:2012uz} the error from the small $R$ approximation remains below $5\%$ for $R=0.7$.} For $pp$ collisions, we have 
\begin{align} \label{eq:fact}
  \frac{\df \si}{\df \eta\, \df p_T\, \df z_r}
  = \sum_{a,b,c} f_a(x_a,\mu)\otimes f_b(x_b,\mu)\otimes {\cal H}_{ab}^c\left(x_a, x_b, \eta, p_T/z, \mu\right)\otimes \cG_c^{\rm jet}(z, z_r, \w_R, \mu)
\,.\end{align}
Here $f_{a,b}$ denote the parton distribution functions and ${\cal H}_{ab}^c$ are hard functions describing the production of an energetic parton of flavor $c$ with transverse momentum $p_T/z$ and rapidity $\eta$ with respect to the beam axis. The subjet functions $\cG_c^{\rm jet}$ describe the subsequent conversion of that parton into a jet moving in (roughly) the same direction but with transverse momentum $z \times p_T/ z = p_T$, containing a subjet of radius $r$ with fraction $z_r$ of the jet energy. The argument $\w_R=2p_T/\cosh \eta$ of $\cG_c^{\rm jet}$ is the large light-cone component of the jet momentum, and the arguments $r$ and $R$ are suppressed. The symbols $\otimes$ denote convolution products associated with the variables $x_{a,b}$ and $z$, which are explicitly written out in~\eq{factorization-pp}. Power corrections to the factorized cross sections are order $R^2$ suppressed. 

We will consider both $r \lesssim R$ and $r \ll R$. In the first case, only single logarithms of the form $\alpha_s^n\ln^n R$ need to be resummed to all orders. The subjet functions $\cG_c^{\rm jet}$ follow timelike DGLAP evolution equations allowing for the resummation of logarithms in the jet size parameter $R$~\cite{Kang:2016mcy,Dai:2016hzf,Dai:2017dpc} (see refs.~\cite{Dasgupta:2014yra,Dasgupta:2016bnd} for a generating functional approach to jet radius resummation). For all subjet observables considered in this work, the resummation of the logarithms of $R$ is the same. For $r \ll R$, we encounter additional large logarithms of $r/R$. The structure and resummation for this class of logarithms depends on how the subjet of size $r$ is identified. For an inclusive subjet sample, we perform an additional collinear factorization for the subjet, matching the subjet functions $\cG_c^{\rm jet}$ onto a semi-inclusive jet function~\cite{Kang:2016mcy,Dai:2016hzf} \emph{for the subjet}. This enables us to resum single logarithms $\as^n\ln^n(r/R)$ using another DGLAP type evolution equation. 

The refactorization for central subjets (i.e.~those centered around a specific axis) in the limit $r\ll R$ differs from the inclusive subjet case, and crucially depends on the choice of axis. The standard jet axis is sensitive to soft radiation inside the jet, since the jet axis is aligned with the total jet momentum. By contrast, the winner-take-all axis is insensitive to soft radiation, but the location of the axis depends on the details of the collinear radiation. For the winner-take-all axis our factorization enables resummation to all-orders in perturbation theory using a (modified) DGLAP evolution~\cite{Neill:2016vbi}, whereas for the standard jet axis this is complicated due to non-global logarithms. 

We will also calculate the average $z_r$ value from \eq{fact} for the central subjet, which corresponds to the jet shape. Its cross section has a single logarithmic dependence on $r/R$ for the winner-take-all axis and a double logarithmic dependence for the standard jet axis. Our factorization formula for the standard jet shape for $r \ll R$ differs from earlier approaches~\cite{Seymour:1997kj,Li:2011hy,Chien:2014nsa}\footnote{From reading ref.~\cite{Chien:2014nsa} one may get the impression that they use a recoil-free axis. The authors confirm that this is not the case.}.
Specifically, it involves a further refactorization to account for the soft radiation that recoils against the jet axis. The additional logarithms of $r/R$ that we can resum, enter the cross section at next-to-leading logarithmic order. This is similar to the broadening event shape where the recoil of soft emissions in the direction of collinear particles (which only enters at NLL order) was initially overlooked~\cite{Catani:1992jc} and only realized later~\cite{Dokshitzer:1998kz}. Like in ref.~\cite{Larkoski:2014uqa}, the effect of recoil can be removed by using the winner-take-all axis. However, in this case this removes all soft sensitivity.

The outline of our paper is as follows: In \sec{cone} we revisit the calculation of the semi-inclusive jet function, addressing an inconsistency in the literature for cone algorithms, and presenting corrected analytical results. We discuss the inclusive production of subjets in \sec{subjetfunction} in terms of the subjet function, for both cone and \kt algorithms, and show numerical results for \eq{obs} for $pp \to (\mathrm{jet}\, j_r)+X$ at NLO+LL$_{R}$+LL$_{r/R}$. In \secs{central_wta}{central} we focus on subjets centered on the winner-take-all axis and the standard jet axis, respectively. In all sections, $r\lesssim R$ as well as $r \ll R$ are considered, and all matching coefficients are calculated at NLO. The jet shape is the second moment (average $z_r$) of the result in \secs{central_wta}{central}, and can be directly related to TMD fragmentation, as discussed in \sec{jetshape}. We conclude in \sec{conclusions} and provide an outlook.

\section{Inclusive cone jets revisited}
\label{sec:cone}

In this section we review the calculation of the semi-inclusive jet functions (siJFs), which enter in the cross section for single inclusive jet production, $pp\to {\rm jet}+X$. Specifically, the cross section for inclusive jet production satisfies the factorization theorem in \eq{fact}, after replacing ${\cal G}_c^{\text{jet}}$ by the siJF $J_c$~\cite{Kang:2016mcy}. We first address an inconsistency in the literature for cone algorithms, before considering the calculation of subjet functions in the following sections (as we also present results for cone algorithms there). Our default notation will be for $e^+e^-$ algorithms, where a jet is defined in terms of its energy $E=\w_R/2$ and angle $R$. These results equally apply to $pp$ algorithms, with the replacement $\w_R R \to 2p_T R$ in terms of a jet radius defined in $(\eta,\phi)$ coordinates, see e.g.~ref.~\cite{Mukherjee:2012uz}.

\begin{figure}[t]\centering
\includegraphics[width=0.6\textwidth]{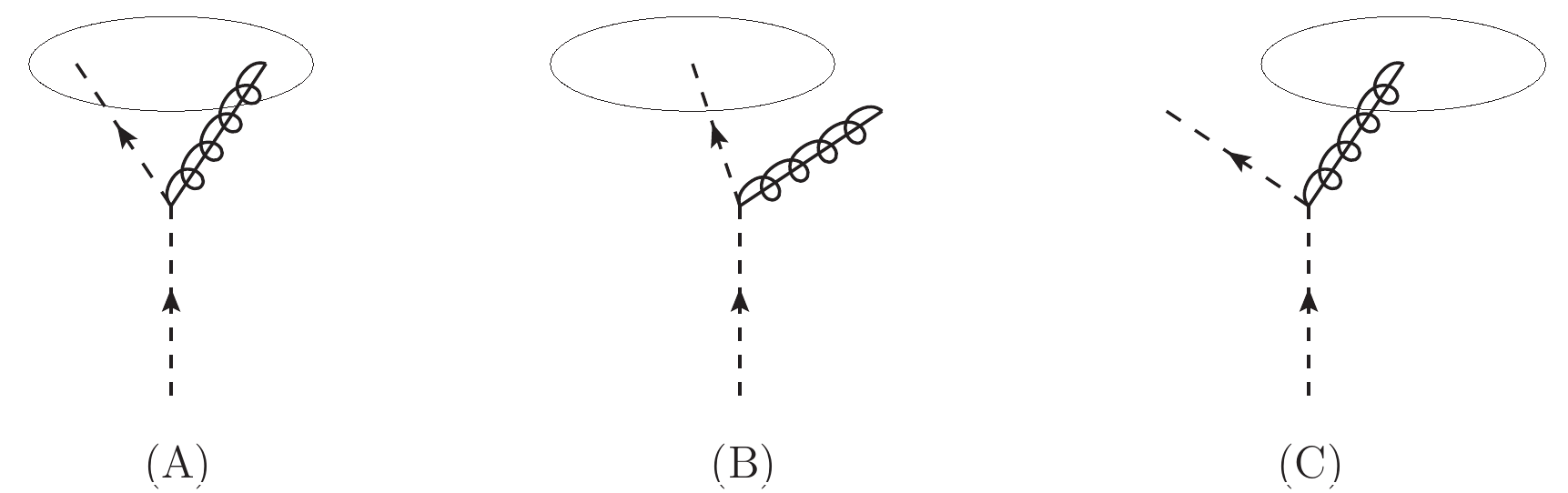}
\caption{The three configurations that enter for the quark semi-inclusive jet function at ${\cal O}(\as)$: (A) the quark and gluon are inside the jet, (B) only the quark is inside the jet, (C) only the gluon is inside the jet.}
\label{fig:inclusivejet}
\end{figure}

For single inclusive jet production in proton-proton collisions at NLO, $pp\to {\rm jet}+X$, there are either one or two final-state partons inside the observed jet, whose possible assignments are illustrated in \fig{inclusivejet}. In refs.~\cite{Aversa:1988vb,Aversa:1989xw,Jager:2004jh,Kaufmann:2015hma,Kang:2016mcy,Kang:2016ehg} the cone algorithm was (effectively) implemented for two final-state partons as 
\begin{align}  
  &\text{Partons in single jet: }& &\beta_1 < R \text{ and } \beta_2 < R\,,
  \nnu
  &\text{Partons in separate jets: }& &\beta = \beta_1 + \beta_2 > R
\,,\end{align}
where $\beta_1$ and $\beta_2$ are the angles of the final state partons with respect to the initiating parton. 
However, these regions of phase space are not complementary, and there are configurations with $R < \beta < 2R$ that are double counted. The resolution depends on the specifics of the cone algorithm. For example, if only the particles themselves are used as seeds for the cone algorithm, the first criterion requires modification and the resulting algorithm happens to coincide with \kt (for two parton configurations). 
For the midpoint~\cite{Akers:1994wj} and the SISCone~\cite{Salam:2007xv} algorithms, the correct implementation is
\begin{align} \label{eq:newconereq}
  &\text{Partons in single jet: }& &\beta_1 < R \text{ and } \beta_2 < R\,,
  \nnu
  & \text{Partons in separate jets: }& &\beta_1 >R \text{ or } \beta_2 > R
\,.\end{align}
Of course the midpoint and the SISCone algorithms will differ with additional particles. 

We now consider the calculation of the semi-inclusive quark jet function. The requirement that the partons are in separate jets in \eq{newconereq}, leads to the following expression for the case where the quark is inside the observed cone jet in \fig{inclusivejet}(B),
\be\label{eq:calccone}
\int \df \Phi_2\,\sigma^c_{2,q}\, \Big[\theta(x<1/2) \theta(\beta_1>R) + \theta(x>1/2) \theta(\beta_2>R)\Big]
\delta(x-z) \,.
\ee
The corresponding expression when the gluon makes the observed jet, \fig{inclusivejet}(C), is obtained by substituting $\delta(x-z)\to\delta(1-x-z)$. The collinear phase space and (squared) matrix element in \eq{calccone} are given by
\begin{align} \label{eq:coll}
\int\! \df \Phi_2\, \si_{2,q}^c = 
\frac{\al_s}{\pi} \f{(e^{\gamma_E}\mu^2)^\epsilon}{\Gamma(1-\epsilon)}\int_0^1 \df x\,
C_F\bigg[\frac{1+x^2}{1-x} - \epsilon\, (1-x) \bigg]
\int\f{\df q_\perp}{q_\perp^{1+2\epsilon}}
\,,\end{align}
where $x$ is the momentum fraction and $q_\perp$ is the transverse momentum of (one of) the final partons with respect to the initiating quark. The angles $\bt_1$ and $\bt_2$ can be expressed in terms of $x$ and $q_\perp$ as follows
\begin{align} \label{eq:beta12}
  \bt_1 = \frac{2 q_\perp}{x \w}
  \,, \qquad
  \bt_2 = \frac{2 q_\perp}{(1-x) \w}
\,,\end{align}
for the partons with momentum fraction $x$ and $(1-x)$. The angle between the partons is
\begin{align} \label{eq:beta}
\bt = \bt_1 + \bt_2 = \frac{2 q_\perp}{x(1-x)\w}
\,.\end{align}

After evaluating the integrals in eq.~(\ref{eq:calccone}) and combining it with the result when both partons are in the jet~\cite{Jager:2004jh,Ellis:2010rwa,Kaufmann:2015hma,Kang:2016mcy}, we obtain the new results for the cone semi-inclusive jet function:
\begin{align} \label{eq:Jq_cone}
J^{\mathrm{cone}}_{q}(z,\omega_R,\mu)  
& = \delta(1-z)+ \f{\as}{2\pi} \bigg\{L_{R} \big[P_{qq}(z)+P_{gq}(z)\big]- 2C_F(1+z^2)\left(\f{\ln(1\!-\!z)}{1-z} \right)_+-C_F
\nnu & \quad
- 2P_{gq}(z)\ln(1-z)+C_F\left(\f72 +3\ln 2-\f{\pi^2}{3}\right)\delta(1-z)
\nnu & \quad 
+ 2\big[P_{qq}(z)+P_{gq}(z)\big]\bigg[\theta\Big(z>\f12\Big)\ln z+\theta\Big(z<\f12\Big)\ln (1-z) \bigg] \bigg\}\, ,
\end{align}
where $L_R$ is defined as
\be
\label{eq:LR}
L_R = \ln\Big(\f{4\mu^2}{\omega_R^2 R^2} \Big) \,.
\ee
The result for gluon-initiated jets can be obtained in a similar way,
\begin{align} \label{eq:Jg_cone}
J^{\mathrm{cone}}_{g}(z,\omega_R,\mu)  
& = \delta(1\!-\!z)+ \f{\as}{2\pi} \bigg\{L_{R} \big[P_{gg}(z)+2n_fP_{qg}(z)\big]- 4C_A\f{(1\!-\!z\!+\!z^2)^2}{z}\left(\f{\ln(1\!-\!z)}{1-z} \right)_+
\nnu & \quad
-4n_f\left[P_{qg}(z)\ln(1-z)+T_Fz(1-z)\right]
\nnu & \quad
+\delta(1-z)\left[C_A\left(\f{137}{36}+\f{11}{3}\ln 2-\f{\pi^2}{3}\right) - T_F n_f\left(\f{23}{18}+\f43\ln 2 \right) \right]
\nnu & \quad 
+ 2\big[P_{gg}(z)+ 2 n_fP_{qg}(z)\big]\bigg[\theta\Big(z>\f12\Big)\ln z+\theta\Big(z<\f12\Big)\ln (1\!-\!z) \bigg] \bigg\}\, . 
\end{align}
In other words,
\begin{align}
J^{\mathrm{cone}}_{q}(z,\omega_R,\mu)  =& J^{\text{cone ref.~\cite{Kang:2016mcy}}}_{q}(z,\omega_R,\mu) 
\nonumber\\
&+ \f{\as}{2\pi} 2\big[P_{qq}(z)+P_{gq}(z)\big]\bigg[\theta\Big(z>\f12\Big)\ln z+\theta\Big(z<\f12\Big)\ln (1-z) \bigg],
\nnu
J^{\mathrm{cone}}_{g}(z,\omega_R,\mu)  =& J^{\text{cone ref.~\cite{Kang:2016mcy}}}_{g}(z,\omega_R,\mu) 
\nonumber\\
&+ \f{\as}{2\pi} 2\big[P_{gg}(z)+ 2 n_fP_{qg}(z)\big]\bigg[\theta\Big(z>\f12\Big)\ln z+\theta\Big(z<\f12\Big)\ln (1\!-\!z) \bigg].
\label{eq:newcone}
\end{align}
As the results for the semi-inclusive jet functions in ref.~\cite{Kang:2016mcy} are consistent with earlier analytical results for single inclusive jet production for cone algorithms in refs.~\cite{Aversa:1988vb,Aversa:1989xw,Jager:2004jh,Kaufmann:2015hma}, the above equations also provide a correction to these earlier cone jet results. As a consistency check, we note that only the updated cone jet results in \eq{newcone} satisfy the following momentum sum rule introduced in~\cite{Dai:2016hzf}\footnote{Ref.~\cite{Dai:2016hzf} only considered \kt jets, and thus did not verify the momentum sum rule for cone jets.}
\be \label{eq:sumrule}
\int_0^1 \df z \, z\, J_i(z,z \omega,\mu)=1 \,,
\ee
where the large momentum component of the \emph{initiating} parton $\omega=\omega_R/z$ is held fixed. Similarly, the NLO matching coefficients for the jet fragmentation function as presented in~\cite{Kaufmann:2015hma,Kang:2016ehg} for cone jets are modified as follows
\begin{align}
{\cal J}^{\text{cone},\one}_{ij}(z,z_h,\omega_R,\mu)= &\; {\cal J}^{\text{cone ref.~\cite{Kang:2016ehg}},\one}_{ij}(z,z_h,\omega_R,\mu) 
\\
&+ \delta(1-z_h)\, \f{\as}{2\pi}\, 2 P_{ji}(z) \left[\theta\Big(z>\f12\Big)\ln z+\theta\Big(z<\f12\Big)\ln (1-z) \right].
\nonumber
\end{align}
For more details, we refer the interested reader to the earlier publications listed above.

\section{Inclusive subjets}
\label{sec:subjetfunction}

In this section, we study the (semi-inclusive) subjet function (SJF), which describes the energy distribution of all subjets inside a jet as in \eq{fact}. 
We gives its definition in \sec{def}, and calculate it to next-to-leading order (NLO) in \sec{NLO}. 
In \sec{RGE} we derive the renormalization group equation (RGE) of the subjet function, which we use to resum the logarithms of the jet radius $R$.
We subsequently consider $r \ll R$ in \sec{matching}, performing the matching onto semi-inclusive jet functions (\siJF) that describe the subjets of radius $r$, and use this to resum the large logarithms of $r/R$. The limit $r\to R$ is discussed in \sec{rR} and the fragmentation limit $r \to 0$ is considered in \sec{frag}. In \sec{excl} we discuss the subjet function for exclusive jet production. We will drop the adjective ``semi-inclusive" in front of the SJF, since we restrict ourselves to inclusive jet samples everywhere else. In \sec{pheno} we show numerical results for the momentum fraction of subjets in $pp \to (\mathrm{jet}\, j_r)+X$ at NLO+LL$_{R}$+LL$_{r/R}$.

\subsection{Definition of subjet function}
\label{sec:def}

We define the subjet function as a matrix element in Soft-Collinear Effective Theory (SCET)~\cite{Bauer:2000ew, Bauer:2000yr, Bauer:2001ct, Bauer:2001yt}. In our definitions and calculations $e^+e^-$ jet algorithms will be our default, which define a jet in terms of its energy $E=\w_R/2$ and angle $R$, and similarly for the subjet. Our results directly apply to $pp$ algorithms with the replacement $E R \to p_T R$, where the jet radius parameter $R$ now refers to a distance in $(\eta,\phi)$ coordinates.
The subjet functions for quark and gluon-initiated jets $\GG_q^{\mathrm{jet}}$ and $\GG_g^{\mathrm{jet}}$ are defined as 
\bea
\GG_q^{\mathrm{jet}}(z, z_r, \omega_R, \mu) =& 16\pi^3\,\sum_X \frac{1}{2N_c}\, {\rm Tr} \Big[\frac{\bnslash}{2}
\langle 0| \delta(\omega - \bar n\cdot {\mathcal P}) \delta^2({\mathcal P}_\perp) \chi_n(0)  |X\rangle 
\langle X|\bar \chi_n(0) |0\rangle \Big]
\nnu & \times
\sum_{J_R\in X} \de\Big(z- \frac{\omega_R}{\omega}\Big) \sum_{j_r \in J_R} \delta\Big(z_r - \frac{\omega_r}{\omega_R}\Big)
\,,\nnu
\GG_g^{\mathrm{jet}}(z, z_r, \omega_R, \mu) =& 16\pi^3\,\sum_X \frac{-\omega}{(d-2)(N_c^2-1)}\, {\rm Tr} \Big[\frac{\bnslash}{2}
\langle 0| \delta(\omega - \bar n\cdot {\mathcal P}) \delta^2({\mathcal P}_\perp) {\mathcal B}_{n\perp}^{\mu,a}(0)  |X\rangle
\nnu & \times 
\langle X|{\mathcal B}_{n\perp, \mu}^a(0) |0\rangle \Big]
\sum_{J_R\in X} \de\Big(z- \frac{\omega_R}{\omega}\Big) \sum_{j_r \in J_R} \delta\Big(z_r - \frac{\omega_r}{\omega_R}\Big)
\,,\eea
suppressing the dependence on $r$ and $R$ in the arguments. 
Here $n^\mu=(1, \hat n)$ is a light-cone vector with its spatial component $\hat n$ along the jet axis, while $\bar n^\mu=(1, -\hat n)$ is a conjugate light-cone vector such that $n^2=\bar n^2=0$ and $n\cdot \bar n = 2$.
$\chi_n$ and ${\mathcal B}_{n\perp}^{\mu}$ are gauge invariant collinear quark and gluon fields in SCET. 
The sum over states $|X\rangle$ runs over all final-state particles and includes their phase-space integrals.
We sum over all reconstructed jets with radius parameter $R$ in $X$ and all subjets $j_r$ with radius $r$. The large light-cone momentum (approximately twice the energy) of the initiating parton, jet and subjet are denoted by $\w$, $\w_R$ and $\w_r$, respectively. For the field producing the initiating parton this is encoded using the (label) momentum operator $\mathcal P$. The variables $z$ and $z_r$ describe the momentum fraction of the initiating parton carried by the jet, and that of the jet carried by the subjet, and are thus given by
\begin{align}
  z = \frac{\w_R}{\w}
  \,, \qquad
  z_r = \frac{\w_r}{\w_R}
\,.  \end{align}

\subsection{NLO calculation}
\label{sec:NLO}

We now calculate the subjet function for quark-initiated jets, $\GG_{q}^{\rm jet}(z,z_r,\omega_R,\mu)$.  For definiteness, we discuss the case where the \kt~algorithm is used for reconstructing both the larger jet of size $R$ and the subjet of size $r$, i.e. we consider ``\kt-in-\kt''. However, at the end of this section we also present results for ``cone-in-cone'' and ``cone-in-\kt'', see \eq{SJF_other}. For ``\kt-in-\kt'' and ``cone-in-cone'' our calculations reveal that, at least at next-to-leading order, these results can be fully expressed in terms of known quantities by \eq{matching_nlo}. The calculation of the gluon subjet function $\cG_g^{\rm jet}$ follows the same steps. Note that the jet algorithms \kt, $k_T$~\cite{Catani:1993hr,Ellis:1993tq} and Cambridge/Aachen~\cite{Dokshitzer:1997in,Wobisch:1998wt} yield the same results to the order that we are considering.

At leading order, the subjet functions are simply given by
\bea\label{eq:GLO}
\cG_i^{\mathrm{jet},(0)}(z, z_r,\omega_R,\mu) = \delta(1-z)\delta(1-z_r) 
\,, \eea
since the total energy of the initiating parton is transferred to the jet (of size $R$) and subjet (of size $r$). 
We perform the next-to-leading order calculation in pure dimensional regularization, where all virtual diagrams vanish. The collinear matrix element and phase-space were given in eq.~\eqref{eq:coll}. The five possible assignments of the partons over the (sub)jet are shown in \fig{configuration}, which we discuss in turn:

\bef
\includegraphics[width=\textwidth]{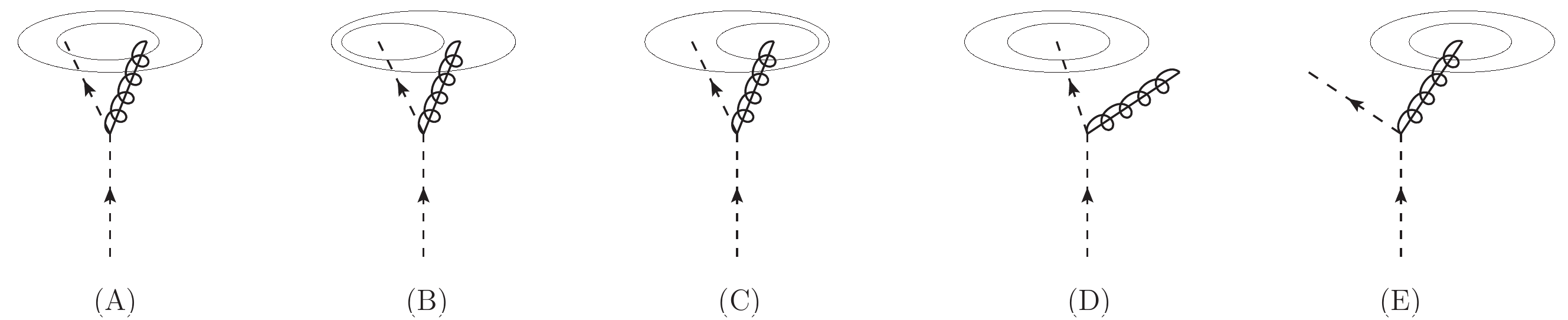}
\caption{The five configurations that enter for the quark subjet function at ${\cal O}(\as)$: (A) the quark and gluon are inside the jet and subjet, (B) only the quark is inside the subjet but both partons are in the jet, (C) only the gluon is inside the subjet but both partons are in the jet, (D) only the quark is inside the jet and subjet, (E) only the gluon is inside the jet and subjet.}
\label{fig:configuration}
\eef

\newpage

\begin{enumerate}[label=(\Alph*)]
\item
\underline{The quark and gluon are inside the jet and subjet}

All the initial quark energy is transferred to the jet and subjet, so $z=\omega_R/\omega=1$ and $z_r=\omega_r/\omega_R=1$. This leads to 
\be\label{eq:J1}
({\rm A}) =
\delta(1-z)\,\delta(1-z_r)
\int\! \df \Phi_2\, \si_{2,q}^c \, \theta(r>\bt)\,,
\ee
where the $\theta$-function encodes the constraint that both partons are inside the jet and subjet when using the \kt~algorithm. Performing the $q_\perp$ and $x$ integrals and expanding in powers of $\epsilon$, we find
\be\label{eq:Jres1}
{\rm (A)}=\delta(1-z)\delta(1-z_r)\,\f{\as C_F}{2\pi}\Big[\f{1}{\eps^2}+\f{3}{2\eps}+\f{L_r}{\eps}+\f{L_r^2}{2}+\f{3}{2}L_r+\f{13}{2}-\f{3\pi^2}{4} + \ord{\eps} \Big],
\ee
where $L_r$ is defined as 
\be
L_r=\ln\left(\f{4\mu^2}{\omega_R^2 r^2}\right) \, .
\ee
We choose to write the results for all configurations (A)\,-\,(E) in terms of $\omega_R$.

\item
\underline{Only the quark is inside the subjet but both partons are inside the jet}

The energy of the quark initiating the jet is transferred entirely to the jet $z=1$, but only the fraction $z_r <1$ is contained inside the subjet. This contribution is
\begin{align}\label{eq:J2}
{\rm (B)} &=
\delta(1-z)
\int\! \df \Phi_2\, \si_{2,q}^c \,\delta(x\!-\!z_r)\, 
\theta(R>\bt>r)
 \\
&=\delta(1\!-\!z)\,\f{\as}{2\pi}\bigg[C_F\,\delta(1\!-\!z_r)\bigg(\!-\!\f{L_{r/R}}{\eps}\!-\!L_R L_{r/R}\!-\!\f{L_{r/R}^2}{2} \bigg) + \f{1+z_r^2}{(1-z_r)}_+\!L_{r/R}
 + \ord{\eps}\bigg].
\nn \end{align}
The $\theta$-function encodes the constraint that the partons are sufficiently close together to be clustered into the jet but not so near that they are in the same subjet. Here $L_R$ is defined in eq.~\eqref{eq:LR} and $L_{r/R}$ is given by
\be
L_{r/R} = L_r - L_R
\,.\ee

\item
\underline{Only the gluon is inside the subjet but both partons are inside the jet}

This configuration is analogous to (B) but exchanging the quark and gluon as shown in fig.~\ref{fig:configuration}(C). This amounts to replacing $\delta(x - z_r) \to \de(1-x - z_r)$ in \eq{J2}, so
\be\label{eq:J3}
{\rm (C)} = \delta(1-z)\,\f{\as}{2\pi}\, L_{r/R}\, P_{gq}(z_r)+ \ord{\eps}
\,,\ee
where the quark splitting functions we use are defined as
\be
P_{qq}(z) = C_F\,\Big(\f{1+z^2}{1-z}\Big)_+ \,,
\qquad
P_{gq}(z) = C_F\,\f{1+(1-z)^2}{z}
\,.
\ee

\item
\underline{Only the quark is inside the jet and subjet}

For the configuration shown in fig.~\ref{fig:configuration}(D), only a fraction $z<1$ of the initiating quark's energy is transferred to the jet of size $R$. However, all the energy of the jet is inside the subjet, $z_r=1$. Thus its contribution to the SJF is given by 
\begin{align}\label{eq:J4}
{\rm (D)} &= \delta(1-z_r) \int\! \df \Phi_2\, \si_{2,q}^c \,\delta(x - z)\, 
\theta(\bt>R)
\nn \\
& = \delta(1-z_r)\,\f{\as C_F}{2\pi} \bigg[\Big(\f{1}{\eps}+L_{R}\Big)\f{1+z^2}{(1-z)}_++ \delta(1-z)\Big(-\f{1}{\eps^2}-\f{L_{R}}{\eps}-\f{L_{R}^2}{2}+\f{\pi^2}{12} \Big) 
\nn \\ & \quad
 -2(1+z^2)\Big(\f{\ln(1-z)}{1-z}\Big)_+ -(1-z) +\ord{\eps}\bigg] \, .
\end{align}
The only constraint from the jet algorithm  is that both partons are far enough apart that they  are not clustered together into the jet.

\item
\underline{Only the gluon is inside the jet and subjet}

This is analogous to (D) but with the replacement $\delta(x - z) \to \delta(1 - x - z)$ in \eq{J4},
\begin{align}\label{eq:J5}
{\rm (E)}
& =\delta(1\!-\!z_r)\,\f{\as}{2\pi}\bigg[\Big(\f{1}{\eps}+L_{R}\Big) P_{gq}(z)- 2 \ln(1 - z) P_{gq}(z) - C_F z + \ord{\eps}\bigg]
.\end{align}

\end{enumerate}

Summing up the leading-order result as well as the five contributions at ${\cal O}(\as)$, we obtain
\begin{align}\label{eq:fixedorder}
\cG^{\mathrm{jet}}_{q,\mathrm{bare}}(z,z_r,\omega_R,\mu)  &=  \cG^{\mathrm{jet},(0)}_q(z,z_r,\omega_R,\mu) + {\rm(A)} + {\rm(B)} + {\rm(C)} + {\rm(D)} + {\rm(E)}
\nn \\& 
  = \delta(1-z)\delta(1-z_r) + \f{\as}{2\pi} \bigg\{\delta(1-z_r)\Big(\f{1}{\eps}+L_{R} \Big)
\big[P_{qq}(z)+P_{gq}(z)\big]
\nnu & \quad
+ \delta(1\!-\!z) L_{r/R}\, [P_{qq}(z_r)+P_{gq}(z_r)] 
+ C_F\,
\delta(1\!-\!z_r)
\bigg[\delta(1\!-\!z)  \Big(\f{13}{2}\!-\!\f{2\pi^2}{3}\Big) 
\nnu & \quad 
- 2(1+z^2)\left(\f{\ln(1\!-\!z)}{1-z} \right)_+ 
- 2\ln(1\!-\!z)\, \frac{1+(1-z)^2}{z} - 1
 \bigg] \bigg\}\, . 
\end{align}
All $1/\eps^2$ poles cancel in the sum as well as all double logarithms $L_{R}^2$ and $L_{r/R}^2$. The result at NLO always involves $\delta(1-z)$ and/or $\delta(1-z_r)$, since the final state consists at most of two partons, but this structure does not generalize to higher orders. The remaining $1/\eps$ pole is a UV divergence that will be removed by renormalization and the resulting time-like DGLAP equation can be used to resum logarithms of $R$, as discussed in the next section. The result for the gluon SJF is:
\begin{align}\label{eq:fixedorderg}
\cG^{\mathrm{jet}}_{g,\mathrm{bare}}(z,z_r,\omega_R,\mu)  &= 
\delta(1-z)\delta(1-z_r) + \f{\as}{2\pi} \bigg\{\delta(1-z_r)\Big(\f{1}{\eps}+L_{R} \Big)
\big[P_{gg}(z)+2 n_f \,P_{qg}(z)\big]
\nnu & \quad
+ \delta(1-z) L_{r/R}\, [P_{gg}(z_r)+2n_f\,P_{qg}(z_r)] 
+ \delta(1-z_r)
\bigg[\delta(1-z)  
\nnu & \quad \times
\bigg(C_A\Big(\f{67}{9}\!-\!\f{2\pi^2}{3}\Big) -T_F n_f\f{23}{9}\bigg)
- 4C_A\f{(1-z+z^2)^2}{z}\left(\f{\ln(1\!-\!z)}{1-z} \right)_+ 
\nnu & \quad 
- 4n_f \Big(P_{qg}(z)\ln(1\!-\!z)\, + T_F z(1-z)\Big)
 \bigg] \bigg\}\,,
\end{align}
which involves the splitting functions
\begin{align}
   P_{gg}(z) &= 2C_A \bigg[ \frac{z}{(1-z)}_+ + \frac{1-z}{z}+z(1-z)\bigg] + \frac{\beta_0}{2}\,\de(1-z)\,,
   \nnu
   P_{qg}(z) &= T_F \big[z^2+(1-z)^2\big]
\,.\end{align}
These results agree with the general setup of ref.~\cite{Dai:2016hzf}. However, there the contributions from (A), (B), (C) are treated separately (factorized) from (D) and (E) (see e.g.~their eq.~(44)). This allows them to relate their expression to exclusive results. As these contributions are at the same scale, the factorization probably fails at higher order in $\alpha_s$. 

For the semi-inclusive fragmenting jet function~\cite{Kang:2016ehg}, a similar structure was obtained as in \eqs{fixedorder}{fixedorderg}. However, this case involved an additional IR pole that cancelled in the matching onto the standard  fragmentation functions. For the SJFs, this IR pole is regulated by the size of the subjet $r$ leaving a single logarithmic dependence on the ratio $r/R$, i.e.~$L_{r/R}$. In \sec{matching}, we are going to match the SJF onto a semi-inclusive jet function for the subjet. This will lead to another time-like DGLAP equation that can be used to resum logarithms of $r/R$.

We conclude this section by giving the renormalized one-loop results for the cone-in-cone and cone-in-\kt SJFs
\bea \label{eq:SJF_other}
&{\mathcal{G}}^{\text{cone-in-cone}}_{q}(z,z_r,\omega_R,\mu) 
\nnu
& = \delta(1\!-\!z)\delta(1-z_r)+\f{\as}{2\pi}\bigg\{\delta(1\!-\!z_r)L_R\left[P_{qq}(z)+P_{gq}(z)\right]
+ \delta(1\!-\!z) L_{r/R} [P_{qq}(z_r)+P_{gq}(z_r)]
\nnu & \quad
 +\delta(1-z)\delta(1-z_r)C_F\left(\f72+3\ln 2-\f{\pi^2}{3} \right)
-\delta(1\!-\!z_r)\bigg[2C_F(1+z^2)\Big(\f{\ln(1\!-\!z)}{1-z}\Big)_+
\nnu & \quad
 +2P_{gq}(z)\ln(1\!-\!z) +C_F
-2[P_{qq}(z)+P_{gq}(z)]\bigg(\theta\Big(z>\f12\Big) \ln z + \theta\Big(z<\f12 \Big)\ln(1-z)\bigg) \bigg]
\bigg\} ,
 \nnu
&\cG^{\text{cone-in-cone}}_{g}(z,z_r,\omega_R,\mu)  
\nnu
&= 
\delta(1-z)\delta(1-z_r) + \f{\as}{2\pi} \bigg\{\delta(1-z_r)L_{R}
\big[P_{gg}(z)+2 n_f \,P_{qg}(z)\big]
+ \delta(1\!-\!z) L_{r/R}\, [P_{gg}(z_r)
 \nnu & \quad 
+2n_f\,P_{qg}(z_r)] 
+ 
\delta(1\!-\!z_r)\delta(1\!-\!z)  \bigg[C_A\Big(\f{137}{36}+\f{11}{3}\ln2\!-\!\f{\pi^2}{3}\Big) -T_F n_f\left(\f{23}{18}+\f43 \ln 2\right)\bigg]
\nnu & \quad
- \delta(1-z_r)\bigg[4C_A\f{(1-z+z^2)^2}{z}\Big(\f{\ln(1\!-\!z)}{1-z} \Big)_+ 
+ 4n_f \big(P_{qg}(z)\ln(1\!-\!z)\, + T_F z(1-z)\big)
\nnu & \quad
-2[P_{gg}(z)+2n_f P_{qg}(z)]\bigg(\theta\Big(z>\f12\Big) \ln z + \theta\Big(z<\f12 \Big)\ln(1-z)\bigg)
 \bigg] 
\bigg\} ,
\nnu
&{\mathcal{G}}^{\text{cone-in-\kt}}_{q}(z,z_r,\omega_R,\mu) 
\nnu
& = \delta(1\!-\!z)\delta(1\!-\!z_r)\!+\!\f{\as}{2\pi}\bigg\{\delta(1\!-\!z_r)L_R [P_{qq}(z)\!+\!P_{gq}(z)]
-\delta(1\!-\!z_r)\bigg[2C_F(1\!+\!z^2)\Big(\f{\ln(1\!-\!z)}{1-z}\Big)_+
\nnu & \quad
 +\!2P_{gq}(z)\ln(1\!-\!z) \!+\! C_F\bigg] 
\!+\! \delta(1\!-\!z)[P_{qq}(z_r)\!+\!P_{gq}(z_r)]\bigg[
\theta\Big(z_r \!>\! \max\Big\{\frac{r}{R},\frac12\Big\}\Big)
(L_{r/R}\!+\!2\ln z_r) 
\nnu & \quad
+ \theta\Big(z_r < \min\Big\{1-\frac{r}{R},\frac12\Big\}\Big)
\big(L_{r/R}+2 \ln (1-z_r)\big) \bigg]
\nnu & \quad
 +\delta(1-z)\delta(1-z_r) C_F \bigg[\theta\Big(r<\frac{R}{2}\Big) \bigg(\f72+3\ln 2-\f{\pi^2}{3} \bigg)
 + \theta\Big(r>\frac{R}{2}\Big) \bigg(-\f{1}{2}L_{r/R}^2+\f{3}{2}L_{r/R}
 \nnu & \quad
-2L_{r/R}\ln\left(1\!-\!\f{r}{R}\right)
 +4\text{Li}_2\left(1\!-\!\f{r}{R}\right) +\f12  - \f{2\pi^2}{3} +6\f{r}{R} \bigg)\bigg]
 \bigg\} ,
 \nnu
& {\mathcal{G}}^{\text{cone-in-\kt}}_{g}(z,z_r,\omega_R,\mu) 
 \nnu
& = \delta(1-z)\delta(1-z_r)+\f{\as}{2\pi}\bigg(\delta(1\!-\!z_r)L_R\left[P_{gg}(z)+2 n_f P_{qg}(z)\right]
\nnu & \quad
-\delta(1-z_r)\bigg[4C_A\f{(1-z+z^2)^2}{z}\Big(\f{\ln(1-z)}{1-z}\Big)_+ +4n_f\big(P_{qg}(z)\ln(1-z) 
+T_Fz(1-z)\big)\bigg]
\nnu & \quad
+ \delta(1-z)[P_{gg}(z_r)+2 n_fP_{gq}(z_r)]\bigg[
\theta\Big(z_r > \max\Big\{\frac{r}{R},\frac12\Big\}\Big)
(L_{r/R}+2\ln z_r) 
\nnu & \quad
+ \theta\Big(z_r < \min\Big\{1-\frac{r}{R},\frac12\Big\}\Big)
\big(L_{r/R}+2 \ln (1-z_r)\big) \bigg]
\nnu & \quad
 +\delta(1-z)\delta(1-z_r)\bigg\{\theta\Big(r<\frac{R}{2}\Big)\bigg[C_A\Big(\f{137}{36}+\f{11}{3}\ln2\!-\!\f{\pi^2}{3}\Big) 
 -T_F n_f\Big(\f{23}{18}+\f43 \ln 2\Big)\bigg]
\nnu & \quad
 + \theta\Big(r>\frac{R}{2}\Big) \bigg[
 C_A \bigg(
 -\f{1}{2}L_{r/R}^2-2 L_{r/R}\ln\left(1\!-\!\f{r}{R}\right)
 +4 \text{Li}_2\left(1\!-\!\f{r}{R}\right)\!-\!\f{2\pi^2}{3} \!+\! \f{8r}{R}\!-\!\f{r^2}{R^2}\!+\!\f{4r^3}{9R^3}\bigg)
\nnu & \quad
 +\f{\beta_0}{2}L_{r/R}
+ T_F n_f\left(\f{1}{3}-\f{4r}{R}+\f{2r^2}{R^2}-\f{8r^3}{9R^3}\right) \bigg] \bigg\}
 \bigg).
\eea

\subsection{Renormalization and resummation of $\ln R$}
\label{sec:RGE}

The renormalization and resulting RG equation of the subjet function is identical to that of the semi-inclusive jet function, because the additional measurement of the subjet does not modify the UV behavior. This also follows from consistency, since the SJFs and siJFs can be interchanged in factorization theorems. For completeness we still present the essential equations. The renormalization of the SJF is given by
\be
\cG_{i,\mathrm{bare}}^{\rm jet}(z,z_r,\omega_R,\mu)=\sum_k\int_z^1\f{\df z'}{z'}\,Z_{ik}\Big(\f{z}{z'},\mu\Big)\, \cG_k^{\rm jet}(z',z_r,\omega_R,\mu)
\,,\ee
which leads to the following RG evolution equation
\ba \label{eq:G_evo}
\mu\f{\df}{\df\mu}\,\cG_i^{\rm jet}(z,z_r,\omega_R,\mu)=\sum_k\int_z^1\f{\df z'}{z'}\,\gamma_{ik}\Big(\f{z}{z'},\mu\Big)\,\cG_k^{\rm jet}(z',z_r,\omega_R,\mu) \, ,
\ea
with the anomalous dimension matrix $\gamma_{ij}$. From our NLO calculation, we immediately obtain
\be
\gamma_{ij}(z, \mu) = \f{\as}{\pi} P_{ji}(z) 
\,,\ee
so the SJF satisfies the usual time-like DGLAP evolution equations.

The one-loop renormalized quark SJF is given by
\begin{align} \label{eq:Gq_finite}
\cG^{\mathrm{jet}}_{q}(z,z_r,\omega_R,\mu)  
& = \delta(1-z)\delta(1-z_r) + \f{\as}{2\pi} \bigg\{ \delta(1-z_r) L_{R} 
\big[P_{qq}(z)+P_{gq}(z)\big]
\nnu & \quad
+ \delta(1\!-\!z) L_{r/R}\, [P_{qq}(z_r)+P_{qq}(1\!-\!z_r)] + C_F
\delta(1\!-\!z_r)
\bigg[\delta(1\!-\!z)  \Big(\f{13}{2}\!-\!\f{2\pi^2}{3}\Big) 
\nnu & \quad 
- 2(1+z^2)\left(\f{\ln(1\!-\!z)}{1-z} \right)_+ - 2\ln(1\!-\!z)\, \frac{1+(1-z)^2}{z} - 1
 \bigg] \bigg\}\, . 
\end{align}
This result for $\cG$ can be used when the jet radii of inner and outer jets are comparable, $r \lesssim R$. By evaluating it at the scale
\be \label{eq:mu_R}
  \mu_R \sim \frac{\omega_R R}{2}\sim p_T R
\, ,\ee
and evolving it with \eq{G_evo} to the scale of the hard scattering 
\be
\mu_H \sim \frac{\omega_R}{2} \sim p_T
\,, \ee
the logarithms of $\mu_R/\mu_H \sim R$ are resummed. When $r\ll R$, the logarithms $L_{r/R}$ in $\cG_i^{\rm jet}$ become large and require resummation as well. This is achieved by an additional factorization, which we discuss next. 

\subsection{Matching for $r \ll R$ and resummation of $\ln (r/R)$}
\label{sec:matching}

In the regime $r \ll R$ we have the following matching equation to all orders in $\alpha_s$
\be
\label{eq:matching}
\GG_i^{\rm jet}(z,z_r,\omega_R,r,R,\mu) = \sum_j \int_{z_r}^1 \frac{\df z_r'}{z_r'} {\mathcal J}_{ij}(z,z_r',\omega_R,R,\mu)\, J_j\Big(\frac{z_r}{z_r'},\omega_r,r,\mu\Big) \bigg[1+ \mathcal{O}\bigg(\frac{r^2}{R^2}\bigg)\bigg]\, ,
\ee
where $J_j$ are the semi-inclusive jet functions describing the subjet of size $r$. In this equation we have explicitly shown the dependence on the jet radius $R$ and the subjet radius $r$ in the arguments of the functions, to highlight its structure.

This matching equation is very similar to that for the matching of the semi-inclusive fragmenting jet function onto fragmentation functions. In fact, the matching coefficients $\cJ_{ij}$ are the same, since they are independent of $r$ and we can therefore safely take the fragmentation limit $r \to 0$. We have also verified this through a direct calculation, and therefore do not give these matching coefficients, but refer the reader to ref.~\cite{Kang:2016ehg} for \kt and \sec{cone} for cone algorithms. 

Interestingly, our calculations reveal that there are no $\ord{r^2/R^2}$ power corrections in \eq{matching} at NLO. In fact, for \kt-in-\kt and cone-in-cone the NLO subjet function is fully determined by
\be \label{eq:matching_nlo}
\GG_i^{{\rm jet}\one}(z,z_r,\omega_R,r,R,\mu) = 
{\mathcal J}_{ij}^\one (z,z_r,\omega_R,R,\mu) +  \de(1-z)\, J_j^\one(z_r,\omega_r,r,\mu)
\,.\ee
For cone-in-\kt, this equation does receive corrections when $r>R/2$.

Having performed the factorization in \eq{matching}, we can now resum the additional logarithms of $r/R$ with the help of another DGLAP RG equation. The scale $\mu_R$ in \eq{mu_R}  sets the large logarithms $L_R$ to zero in the matching coefficients $\cJ_{ij}$, whereas 
\be
  \mu_r\sim\frac{\omega_r r}{2}\sim p_T r
\, ,\ee
is the natural scale for semi-inclusive jet function $J_j$ describing the subjet in \eq{matching}. By using the RG equation to evolve the \siJF from $\mu_r$ to $\mu_R$, we resum the single logarithms of $r/R$.

\subsection[The limit $r\to R$]{The limit $r \to R$}
\label{sec:rR}

We will now start from the regime $r\ll R$, for which the logarithms of $r/R$ are resummed, and consider the limit when the inner jet radius becomes large. It is fairly straightforward to do this, because the absence of power corrections in \eq{matching} at this order. The aim is to gain an analytical understanding of the  $r\to R$ limit, for which we consider \kt-in-\kt at leading-logarithmic accuracy.

We start by observing that for $z_r<1$, the only terms in \eq{fixedorder} that contribute are
\be\label{eq:FOcontribution}
{\cal G}_q^{\mathrm{jet}}(z,z_r<1,\omega_R,\mu)= \f{\alpha_s}{2\pi}\,\delta(1-z)L_{r/R}\left[P_{qq}(z_r)+P_{gq}(z_r)\right]
\,,\ee
since all other terms are proportional to $\delta(1-z_r)$. Note that the role of the variables $z$ and $z_r$ are fundamentally different, in the sense that $z$ is an integration variable for the convolution with a hard function and $z_r$ is the measured external variable. As can be seen from \eq{FOcontribution}, the subjet cross section at NLO is directly proportional $\ln(r/R)$ and, therefore, it can be considered as a direct probe of the $\ln(r/R)$ resummation.

In the limit $r\ll R$, we refactorize the subjet function ${\cal G}_i^{\mathrm{jet}}$ in terms of matching coefficients and evolved siJFs, see \eq{matching}. In order to recover the NLO result in \eq{FOcontribution} from the resummed result in the limit $r\to R$, we find that it is sufficient to consider the leading-order matching coefficients
\be\label{eq:matchinglo}
{\cal J}_{ij}^{(0)}(z,z_r,\omega_R,\mu_R)=\delta_{ij}\delta(1-z)\delta(1-z_r) \,,
\ee
as well as the leading-order initial condition for the siJFs for the subjets
\be\label{eq:siJFlo}
J_i^{(0)}(z_r,\omega_r,\mu_r)=\delta(1-z_r)
\,.\ee
Including ${\cal O}(\alpha_s)$ corrections in the matching or initial condition would generate $\ord{\al_s^2}$ terms in ${\cal G}_i^{\mathrm{jet}}$.

Using the techniques of refs.~\cite{Kang:2016mcy,Kang:2016ehg}, we solve the DGLAP equations associated with the resummation of both logarithms $\ln (r/R)$ and $\ln R$ in Mellin moment space. In order to perform the resummation, we take double Mellin moments of the subjet functions ${\cal G}^{\mathrm{jet}}_i$
\be
{\cal G}^{\mathrm{jet}}_i(M,N,\omega_R,\mu)=\int_0^1\df z\, z^{M-1}\int_0^1 \df z_r\, z_r^{N-1}\,{\cal G}_i^{\mathrm{jet}}(z,z_r,\omega_R,\mu) \,.
\ee
We only discuss the resummation of logarithms $\ln(r/R)$ associated with the variable $z_r$ and the Mellin variable $N$. (The DGLAP equation for resumming $\ln R$ was given in \eq{G_evo} and is associated with the variable $z$ and Mellin variable $M$.) The convolution of matching coefficients and the siJFs in \eq{matching} turns into simple products
\be
{\cal G}^{\mathrm{jet}}_i(M,N,\omega_R,\mu)=\sum_j {\mathcal J}_{ij}(M,N,\omega_R,\mu)\, J_j(N,\omega_r,\mu) \,.
\ee
The delta functions of the leading-order matching coefficients in \eq{matchinglo} integrate to 1 when taking moments, so the subjet function in Mellin space is given by the siJF evolved from $\mu_r$ to $\mu_R$. For the quark siJF at LL accuracy, we find 
\be
J_q(N,\omega_r,\mu_R)=\left(\f{\alpha_s(\mu_R)}{\alpha_s(\mu_r)}\right)^{-\f{2}{\beta_0}[P_{qq}(N)+P_{gq}(N)]} \,,
\ee
where $P_{ji}(N)$ are the leading-order Altarelli-Parisi splitting functions in Mellin $N$ moment space and $\beta_0$ is the first coefficient of the QCD beta function. Inserting the leading-order solution for the running strong coupling constant
\be
\f{1}{\alpha_s(\mu_R)}=\f{1}{\alpha_s(\mu_r)}+\f{\beta_0}{2\pi}\ln\left(\f{\mu_R}{\mu_r}\right) \,,
\ee
this becomes
\be
J_q(N,\omega_r,r,\mu_R)= \exp\left[\f{2}{\beta_0}\left(P_{qq}(N)+P_{gq}(N)\right) \ln\left(1+\f{\alpha_s(\mu_r)}{2\pi}\beta_0\ln\left(\f{\mu_R}{\mu_r}\right)\right)\right] \,.
\ee
In the limit $r\to R$,
 \be\label{eq:resMellin}
J_q(N,\omega_r,r,\mu_R) = 1+\f{\alpha_s(\mu_R)}{2\pi} L_{r/R} [P_{qq}(N)+P_{gq}(N)] + \ord{\al_s^2}\,.
 \ee
 
Finally, we need to perform the double Mellin inverse transformation of the whole subjet function which is given by contour integrals in the complex $M$ and $N$ planes
\be\label{eq:MellinInverse}
{\cal G}^{\mathrm{jet}}(z,z_r,\omega_R,\mu)=\int_{{\cal C}_M}\f{\df M}{2\pi \img}\, z^{-M}\int_{{\cal C}_N}\f{\df N}{2\pi \img}\, z_r^{-N}\, {\cal G}^{\mathrm{jet}}(M,N,\omega_R,r,R,\mu) \,.
\ee
The first term in \eq{resMellin} does not contribute to the cross section, since it does not contain poles in $N$. The second term in \eq{resMellin} directly gives the NLO contribution for $z_r<1$ shown in \eq{FOcontribution} above. Therefore, we have recovered the fixed-order cross section from the resummed result in the limit of $r\to R$. We also verified this numerically in \sec{pheno}. Note that the inverse with respect to $N$ in \eq{MellinInverse} can be taken directly as it is associated with the observed external variable $z_r$. However the resulting expression for $z$ still needs to be convolved with the hard function.

\subsection{The fragmentation limit $r\to 0$}
\label{sec:frag}

For sufficiently small $r$, the scale $\w_r r/2$ of the semi-inclusive jet function with radius $r$ becomes nonperturbative. At that point we can no longer speak of subjets but are really probing individual hadrons, and the subjet function should be replaced by a fragmenting jet function (inclusive in hadron species). This limit is continuous, since the matching coefficients are the same whether we match onto subjets or hadrons, as discussed in \sec{matching}.
We stress that our conclusions also apply to inclusive jet cross sections, as we will study the nonperturbative corrections to the semi-inclusive jet function.

We start by factorizing the \siJF into a perturbative and nonperturbative component, 
\bea \label{eq:nonpert}
  J_i(z_r,\w, \mu) = \int\! \frac{\df z'}{z'} J_i^{\rm pert}(z_r', \w, \mu) J_i^{\rm NP}\Big(\frac{z_r}{z_r'}, \w r \Big)
\,.\eea
In contrast to the rest of the paper, we work in terms of the large momentum component $\w$ of the initiating parton $i$, instead of $\w_r = z \w$ of the (sub)jet. $J_i^{\rm pert}$ is the perturbative result (which was calculated at NLO in refs.~\cite{Kang:2016mcy,Dai:2016hzf}) and $J_i^{\rm NP}$ captures the nonperturbative corrections. From boost invariance (or reparametrization invariance~\cite{Manohar:2006nz}) we infer that the arguments $\w$ and $r$ appear in the combination $\w r$, which was exploited in writing the arguments of $J_i^{\rm NP}$. Since $J_i^{\rm pert}$ has the same anomalous dimension as the full $J_i$, this implies that $J_i^{\rm NP}$ is independent of $\mu$. Note that this crucially relies on including the nonperturbative corrections through a Mellin convolution in \eq{nonpert}, since the anomalous dimension is only diagonal in Mellin space. 

Although $J_i^{\rm NP}$ is a two-dimensional nonperturbative function, we know its limits: 
\begin{align} \label{eq:limits}
  &\w r \to \infty:& & J_i^{\rm NP}(z_r, \w r) \to \de(1-z_r)
  \,, \nnu
  &\w r \to \lqcd:& & J_i^{\rm NP}(z_r, \w r) \to \sum_h D_i^h(z_r, \mu = \w r/2)
\,.\end{align}
The first line is the perturbative limit, for which the nonperturative corrections (but not $J_i^{\rm NP}$) vanish. From the continuity of the $r \to 0$ fragmentation limit of subjets, it follows that the semi-inclusive jet function turns into the fragmentation function (summed over hadron species), as shown on the second line.

\begin{figure*}[t]
\centering
\hfill \includegraphics[width=0.45\textwidth]{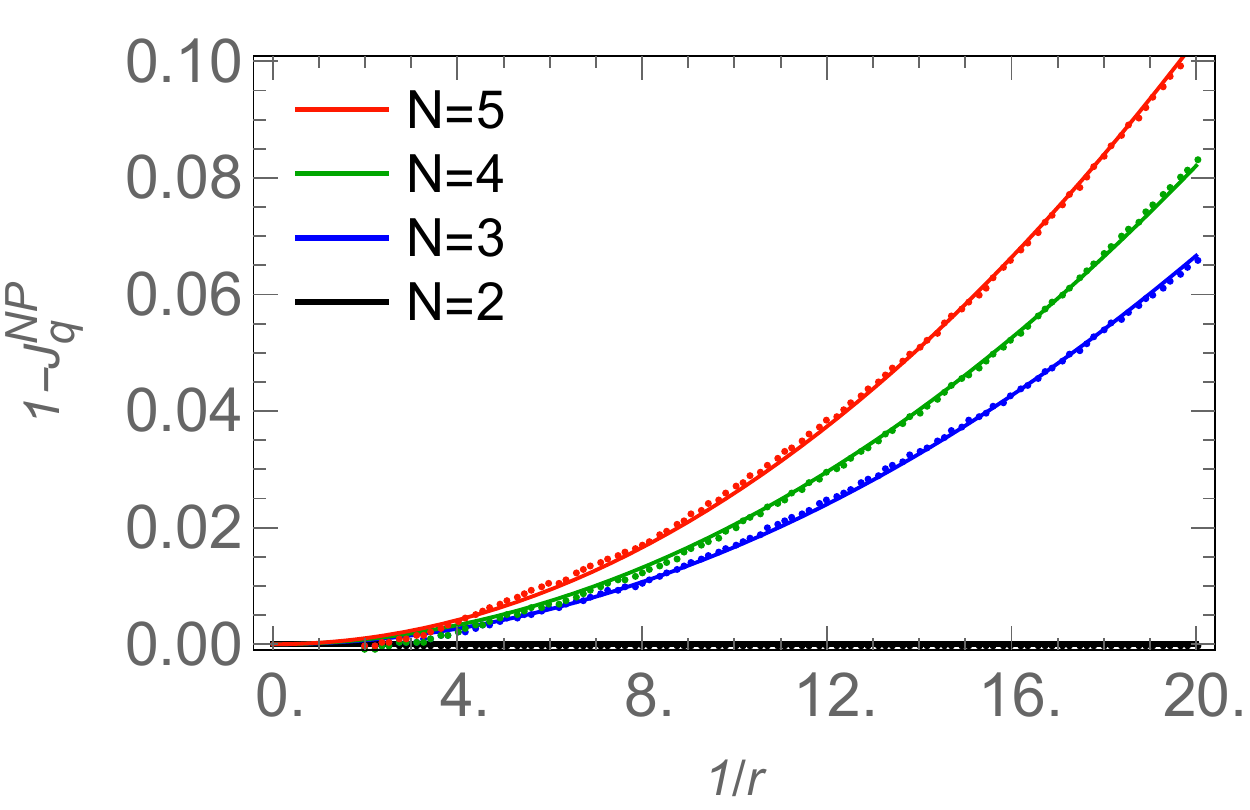} \hfill
\includegraphics[width=0.445\textwidth]{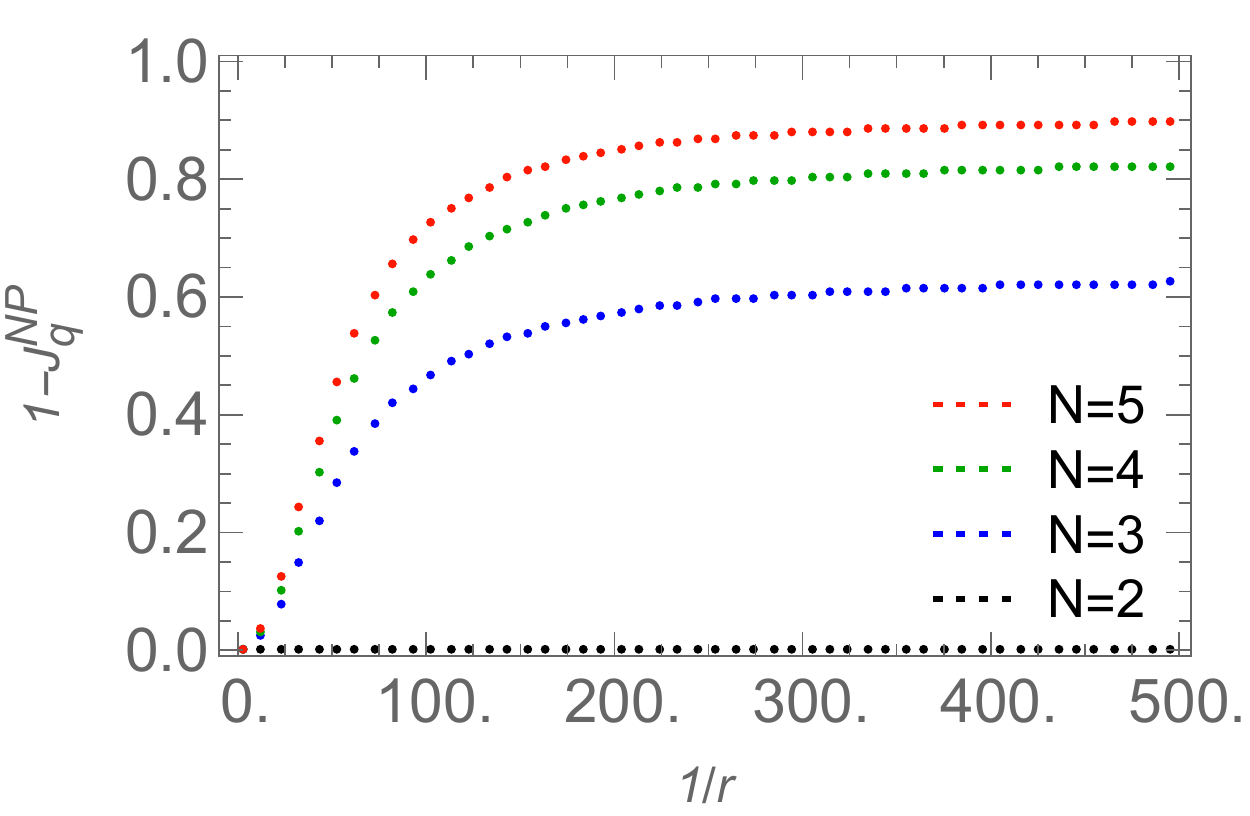} \hfill \hfill 
\caption{\label{fig:nonp} The nonperturbative corrections $1-J_q^{\rm NP}$ to the semi-inclusive quark jet function for Mellin moments $N=2$ (black), 3 (blue), 4 (green), 5 (red) as function of $1/r$. The points were extracted from parton-level and hadron-level \textsc{Pythia} data, using $e^+e^-$ collisions at a center-of-mass energy of $\w= 500$ GeV, with the $e^+e^-$ \kt algorithm. In the left panel we restrict to large values of $r$ and show a fit to $c/r^2$ (solid lines). In the right panel, the asymptotic approach to the fragmentation limit is shown. The nonperturbative corrections for $N=2$ vanish due to \eq{sumrule}.}
\end{figure*}

We have extracted $J_q^{\rm NP}$ from \textsc{Pythia}~\cite{Sjostrand:2014zea}, using parton-level and hadron-level inclusive jet spectra for $e^+e^-$ collisions at a center-of-mass energy of 500 GeV, with the $e^+e^-$ \kt algorithm. This implies that $\w =500$ GeV (whereas $\w_r$ varies). Instead of considering the full $z_r$ dependence we take Mellin moments, such that $J_q^{\rm NP}$ is the ratio of hadron-level and parton-level cross sections. The result is shown in \fig{nonp} as function of $1/r$. In the perturbative limit $\w r \to \infty$,  $J_q^{\rm NP}(N, \w r) \to 1$, so to visualize the nonperturbative effects we plot $1-J_q^{\rm NP}$. Also, $J_q^{\rm NP}(N\!=\!2, \w r) = 1$, due to the momentum sum rule in \eq{sumrule} (which relies on using $\w$ rather than $\w_r$). In the left panel we limit ourselves to $\w r/2 > 10$ GeV, for which it is reasonable to carry out a series expansion in $2\lqcd/(\w r)$. We find that the linear term vanishes and the quadratic term (fit shown as solid line in left panel) describes the points very well. There is a slight discrepancy for large values of $r$, but in this regime the $\ord{r^2}$ corrections to the factorization theorem may no longer be negligible. In the right panel of \fig{nonp} we focus on the nonperturbative regime, displaying the asymptotic behavior. For $\w r/2 \sim 5$ GeV the nonperturbative corrections deviate from the quadratic fit by about 10\%, whereas for $\w r/2 \lesssim 1$ GeV, the jets essentially consist of single hadrons.

\subsection{Subjets in exclusive jets}
\label{sec:excl}

Up to this point we have focussed entirely on inclusive jet production. However, one can also consider exclusive jet production, where additional jets are vetoed. The collinear (energetic) radiation is then forced to be inside the jet. Adding this additional restriction to the definition of the subjet function in \sec{def}, 
\bea
\GG_q^{\mathrm{jet}}(z_r, \omega_R, \mu) =& 16\pi^3\,\sum_{J_R} \frac{1}{2N_c}\, {\rm Tr} \Big[\frac{\bnslash}{2}
\langle 0| \delta(\omega_R - \bar n\cdot {\mathcal P}) \delta^2({\mathcal P}_\perp) \chi_n(0)  |J_R \rangle 
\langle J_R|\bar \chi_n(0) |0\rangle \Big]
\nnu & \times
\sum_{j_r \in J_R} \delta\Big(z_r - \frac{\omega_r}{\omega_R}\Big)
\,.\eea
In this case there is no collinear radiation outside the jet, which is why we replaced $X$ by $J_R$. Consequently, there is no dependence on $z$, since $z=\w_R/\w = 1$ always.

In the NLO calculation, the contributions in \fig{configuration}(D) and (E) are absent. This does not modify the dependence on $z_r$ and $r$, but removes the $z$ dependence and introduces double logarithms of $R$. 
These logarithms of $R$ can again be resummed using the RGE of the subjet function. However, instead of the convolution structure seen in \eq{G_evo}, the RGE of the subjet function is now multiplicative, 
\bea
\mu\f{\df}{\df\mu}\,\cG_{i,{\rm excl}}^{\rm jet}(z_r,\omega_R,\mu) = \gamma_{i,{\rm excl}}(\w_R, R,\mu)\,\cG_{i,{\rm excl}}^{\rm jet}(z_r,\omega_R,\mu) \, ,
\eea
The anomalous dimension $\ga_{i,{\rm excl}}$ is the same as for the unmeasured jet functions of ref.~\cite{Ellis:2010rwa}, and given in eq.~(6.26) therein. The appearance of double logarithms of $R$ indicate a sensitivity to soft radiation and so the factorization in \eq{fact} must be modified to include a soft function.

The matching for $r \ll R$ in \sec{matching} onto semi-inclusive jet functions still holds, 
\be
\GG_{i,{\rm excl}}^{\rm jet}(z_r,\omega_R,r,R,\mu) = \sum_j \int_{z_r}^1 \frac{\df z_r'}{z_r'} {\mathcal J}_{ij,{\rm excl}}(z_r',\omega_R,R,\mu)\, J_j\Big(\frac{z_r}{z_r'},\omega_r,r,\mu\Big) \bigg[1+ \mathcal{O}\bigg(\frac{r^2}{R^2}\bigg)\bigg]\, ,
\ee
but the matching coefficients $\cJ_{ij, {\rm excl}}$ are not the same as in the inclusive case. Rather, they are the same as those of the fragmenting jet functions for exclusive jet samples, which were calculated in ref.~\cite{Procura:2011aq} for cone algorithms and in refs.~\cite{Waalewijn:2012sv,Chien:2015ctp} for \kt.

\subsection{Phenomenology for  $pp\to(\mathrm{jet}\, j_r)+X$}
\label{sec:pheno}

\begin {figure*}[t]
\begin{center}
\vspace*{10mm}
\includegraphics[width=0.4\textwidth,trim=1cm 2cm 1cm 1cm ]{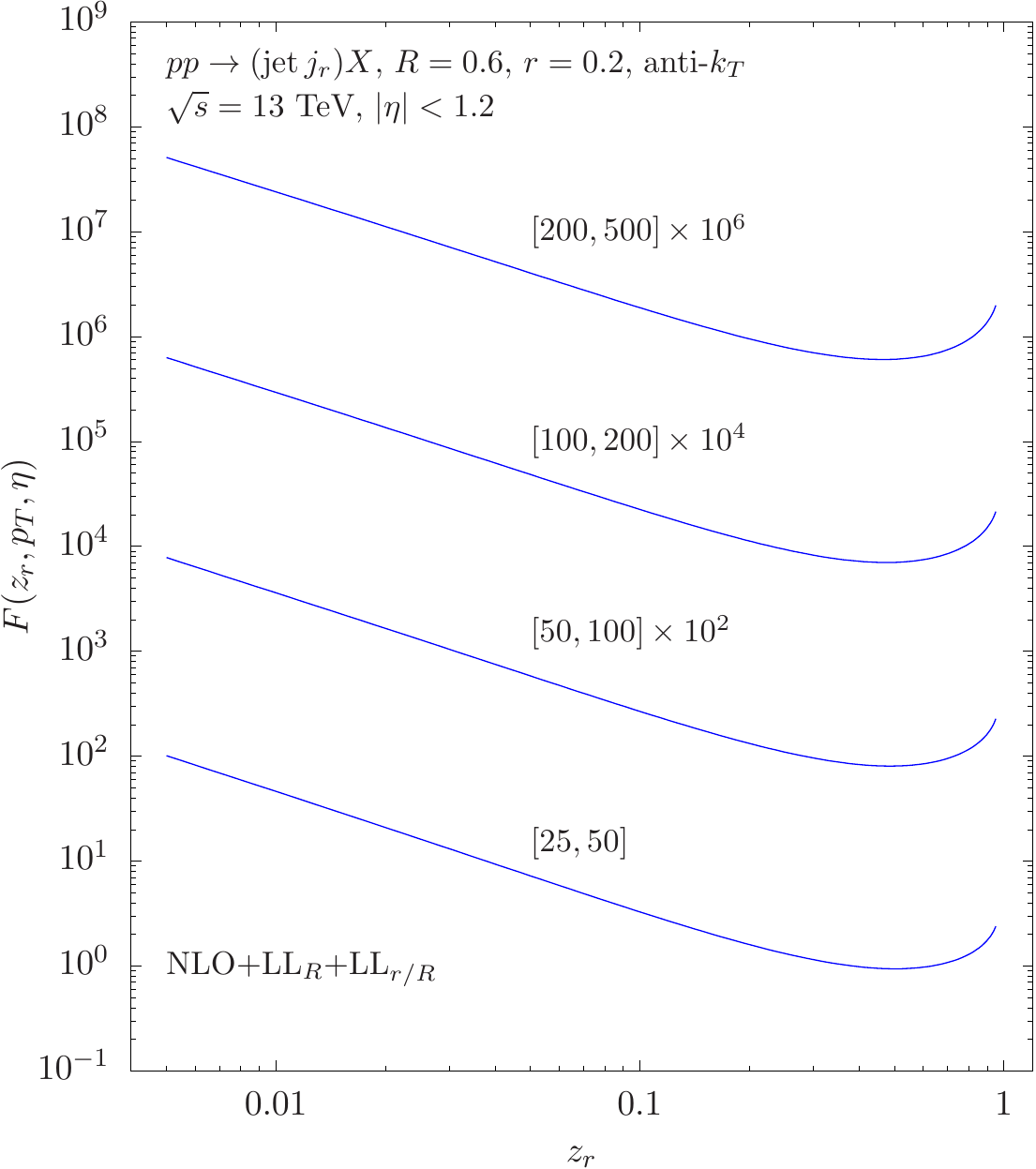} 
\end{center}
\vspace*{1.cm}
\caption{\label{fig:SJF-1} The subjet distribution measured on an inclusive jet sample $pp\to (\mathrm{jet}\, j_r)+X$, using \kt with jet radius $R=0.6$ and subjet radius $r=0.2$, for representative LHC kinematics $\sqrt{s}=13$~TeV, $|\eta|<1.2$. Shown are the NLO+LL$_R$+LL$_{r/R}$ results  for four different intervals of the jet transverse momentum $[25,50],\; [50,100],\; [100,200],\; [200,500]$~GeV.}
\end{figure*}

We present numerical results for the momentum fraction of subjets measured on an inclusive jet sample $pp\to\mathrm{jet}+X$. In analogy with the hadron-in-jet calculations in proton-proton collisions presented in refs.~\cite{Kaufmann:2015hma,Kang:2016ehg}, we adopt the notation $pp\to(\mathrm{jet}\, j_r)+X$, where $j_r$ denotes a subjet of size $r$ inside the larger jet of size $R$. The factorization formula for the subjet distribution in proton-proton collisions is given by
\bea
\label{eq:factorization-pp}
\frac{\df\sigma^{pp\to (\mathrm{jet}\, j_r)X}}{\df p_T\,\df \eta\, \df z_r}  = & \sum_{a,b,c}\int_{x_a^{\mathrm{min}}}^1\!\f{\df x_a}{x_a}\,f_a(x_a,\mu)\int_{x_b^{\mathrm{min}}}^1\!\f{\df x_b}{x_b} f_b(x_b,\mu) 
\nnu
&\times
\int^1_{z^{\mathrm{min}}}\! \frac{\df z}{z^2}\,{\cal H}^c_{ab}(\hat s,\hat p_T,\hat \eta,\mu)\,{\cal G}_c^{\rm jet}(z,z_r,\omega_R,\mu)\,,
\eea
where the sum on $a,b,c$ runs over all relevant partonic channels. The PDFs are denoted by $f_{a,b}$ and the hard functions are given by ${\cal H}_{ab}^c$, which have been calculated to NLO in refs.~\cite{Aversa:1988vb,Jager:2002xm}. For all numerical results presented in this section, we use the CT14 NLO set of PDFs~\cite{Dulat:2015mca}. The variables $s$, $p_T$ and $\eta$ correspond to the center-of-mass (CM) energy, the jet transverse momentum and the jet rapidity respectively. The hard functions depend on the corresponding partonic variables $\hat s=x_ax_bs$, $\hat p_T=p_T/z$ and $\hat\eta=\eta-\ln(x_a/x_b)/2$. The lower integration bounds $x_a^{\mathrm{min}}$, $x_b^{\mathrm{min}}$ and $z^{\mathrm{min}}$ can be written in terms of these variables and are listed for example in refs.~\cite{Kaufmann:2015hma,Kang:2016ehg}. 

The subjet function ${\cal G}_c^{\rm jet}$ in \eq{factorization-pp} is evolved to the hard scale $\mu\sim p_T$ by solving the DGLAP evolution equations associated with the logarithms $\ln R$ and $\ln(r/R)$. Numerically, we solve the DGLAP equations in Mellin moment space using the techniques developed in refs.~\cite{Kang:2016mcy,Kang:2016ehg}, which in turn are based on the evolution packages of refs.~\cite{Vogt:2004ns,Anderle:2015lqa}. We jointly resum both single logarithms $\ln R$ and $\ln(r/R)$ with a combined accuracy of ``NLO+LL$_R$+LL$_{r/R}$''. The evolved subjet function ${\cal G}_c^{\rm jet}$ is divergent for $z\to 1$. We can nevertheless perform the integrals in eq.~(\ref{eq:factorization-pp}) by adopting the prescription of ref.~\cite{Bodwin:2015iua}, as discussed in detail in ref.~\cite{Kang:2016mcy}. Note that the factorized form of the cross section in \eq{factorization-pp} is a purely collinear factorization, i.e.~there is no soft function. All numerical results presented here are normalized by the total inclusive jet cross section, see~\eq{obs}, for which we resum single logarithms of the jet size parameter $\ln R$ at NLO+LL$_R$ accuracy. 

\begin {figure*}[t]
\begin{center}
\vspace*{10mm}
\includegraphics[width=0.4\textwidth,trim=1cm 2cm 1cm 1cm ]{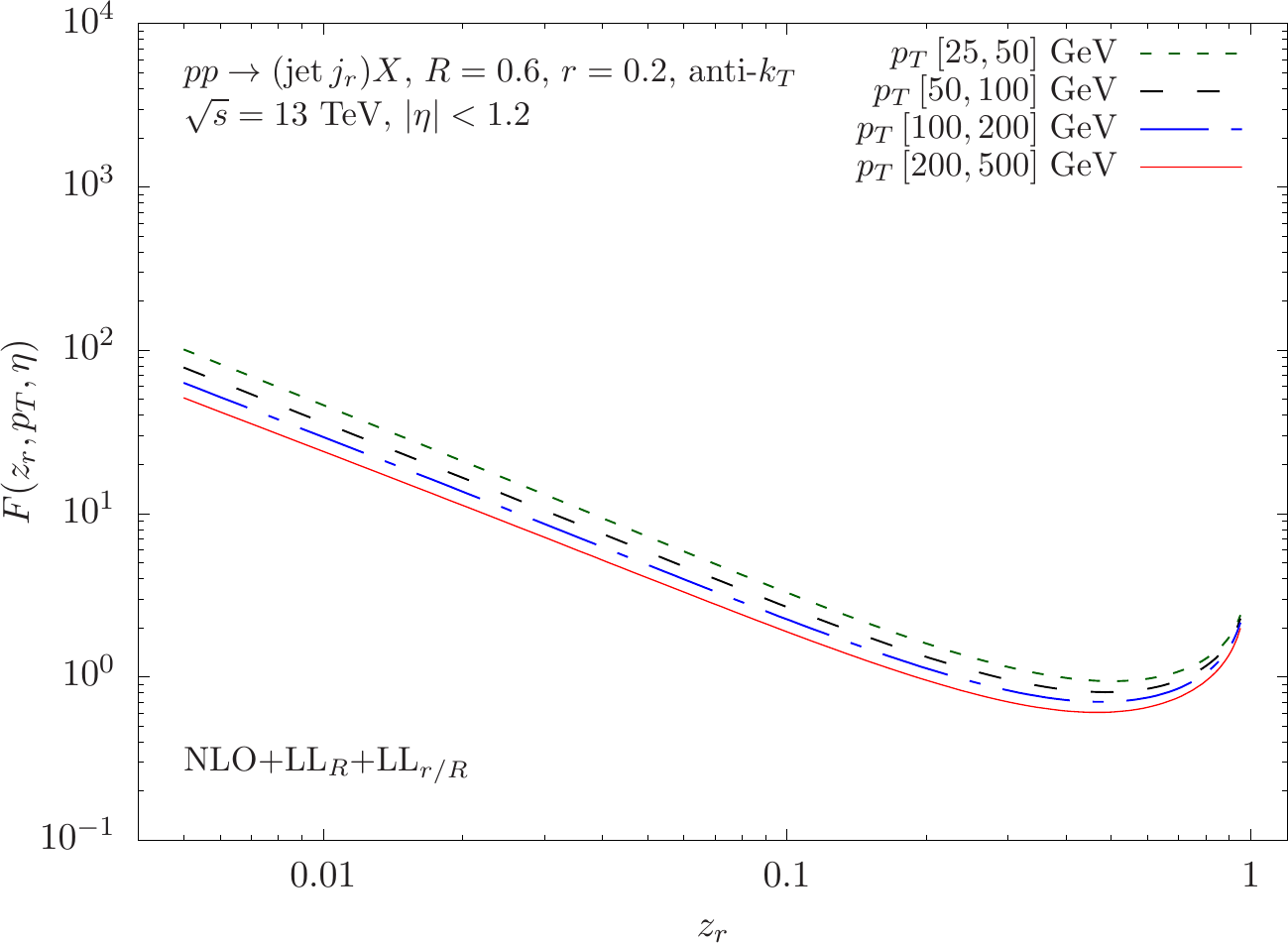} 
\hspace*{2cm}
\includegraphics[width=0.4\textwidth,trim=1cm 2cm 1cm 1cm ]{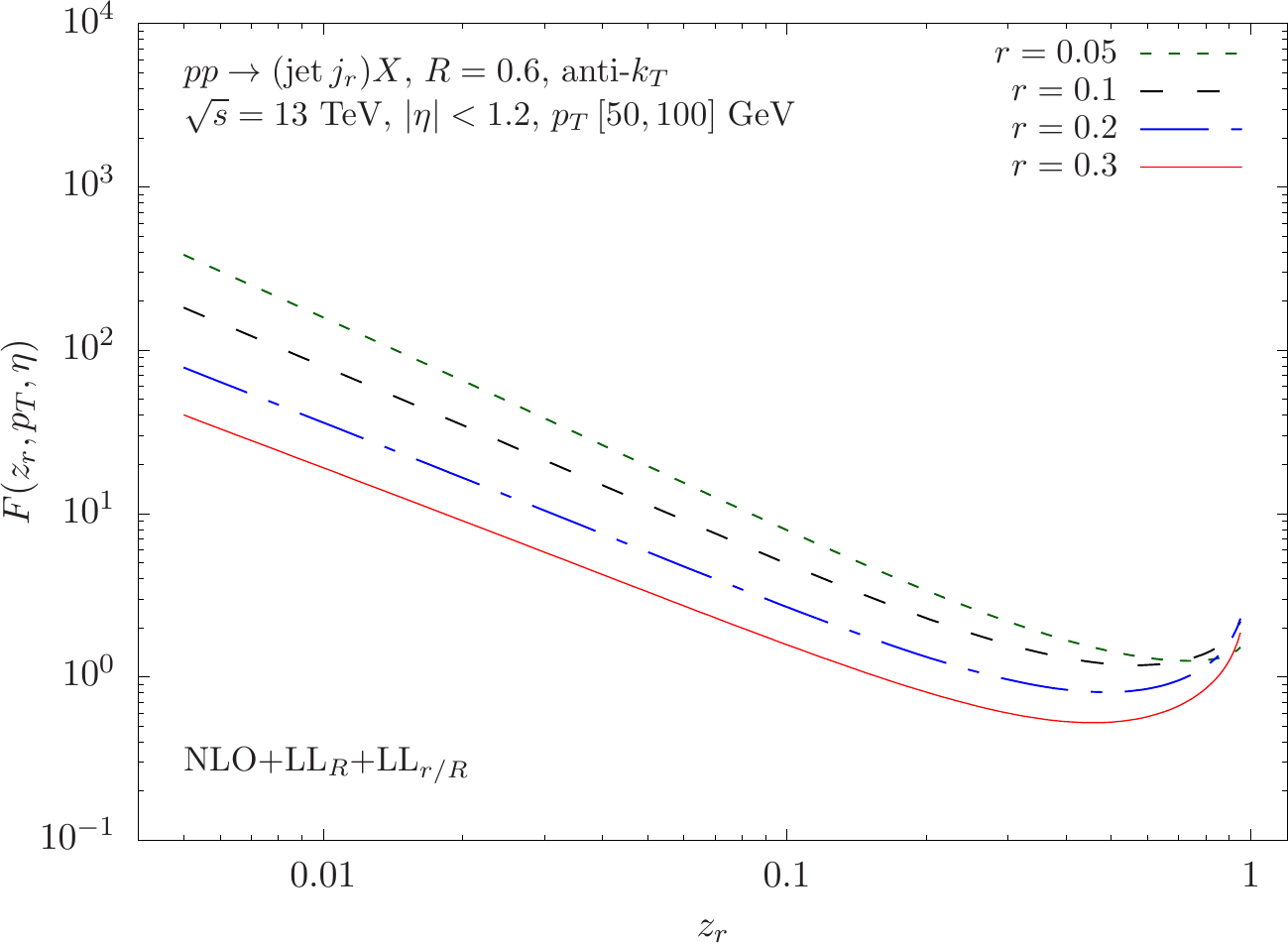}
\end{center}
\vspace*{1.cm}
\caption{\label{fig:SJF-2} Left: Same as \fig{SJF-1} but without multiplying the result for the different $p_T$ intervals by multiples of 10. Right: Subjet distribution measured on an inclusive jet sample, using the same kinematics as in \fig{SJF-1} and the $p_T$ bin of $[50,100]$~GeV. The distribution is shown for different values of the subjet radius: $r=0.05$ (green dotted), $r=0.1$ (black dashed), $r=0.2$ (blue dot-dashed) and $r=0.3$ (red solid).}
\end{figure*}

For our numerical results we choose representative LHC kinematics, taking a CM energy of $\sqrt{s}=13$~TeV and a rapidity range of $|\eta|<1.2$. Both the jet and the subjets are identified using the anti-$k_T$ algorithm. We choose a jet radius of $R=0.6$ for the outside jet. 
In \fig{SJF-1} we plot the momentum fraction $z_r$ for a subjet radius parameter of $r=0.2$ for  different bins of the transverse momentum $p_T$ of the jet. We multiply the results for the different $p_T$ bins by multiples of 10 for better visibility. One immediately notices that the plotted curves look like the QCD Altarelli-Parisi splitting functions. This behavior of the cross section can be most easily understood by looking at the fixed order results for the subjet function in \eq{Gq_finite}. Only the terms  $\ln(r/R) P_{ji}(z_r)$ have a non-trivial functional dependence on $z_r$, as all other terms are proportional to $\delta(1-z_r)$ and do not contribute at fixed order. The $\ln(r/R)$ resummation modifies the distribution slightly, so it is not exactly the splitting function.

We would like to point out an important difference of the results for the subjet distribution compared to the distribution of light charged hadrons inside jets, as presented in for example refs.~\cite{Kaufmann:2015hma,Kang:2016ehg}. When measuring an identified hadron inside jets, the distribution falls continuously as $z_h$ increases. However, as can be seen from fig.~\ref{fig:SJF-1}, the distribution of subjets starts to rise again for sufficiently large $z_r$. Whereas it becomes increasingly unlikely to find a hadron that carries a large fraction of the complete jet,  a subjet with radius $r<R$ may still contain most of the energy of the larger outside jet as long as $r$ is not too small. In order to better see the dependence on the jet $p_T$, we plot on the left panel of \fig{SJF-2}, the same curves as in \fig{SJF-1} but without multiplying them by multiples of 10. We observe only a relatively small dependence on the jet transverse momentum.

\begin {figure*}[t]
\begin{center}
\vspace*{10mm}
\includegraphics[width=0.4\textwidth,trim=1cm 2cm 1cm 1cm ]{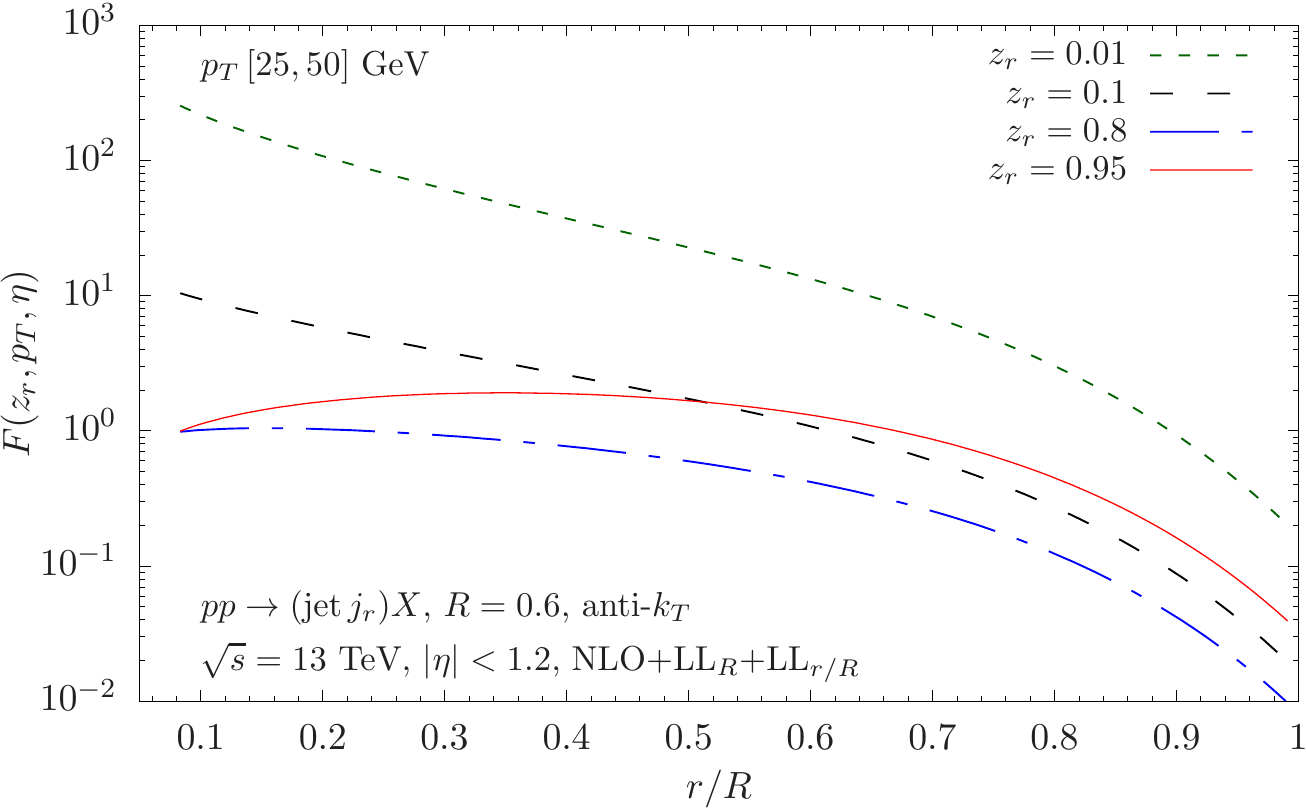} 
\hspace*{2cm}
\includegraphics[width=0.4\textwidth,trim=1cm 2cm 1cm 1cm ]{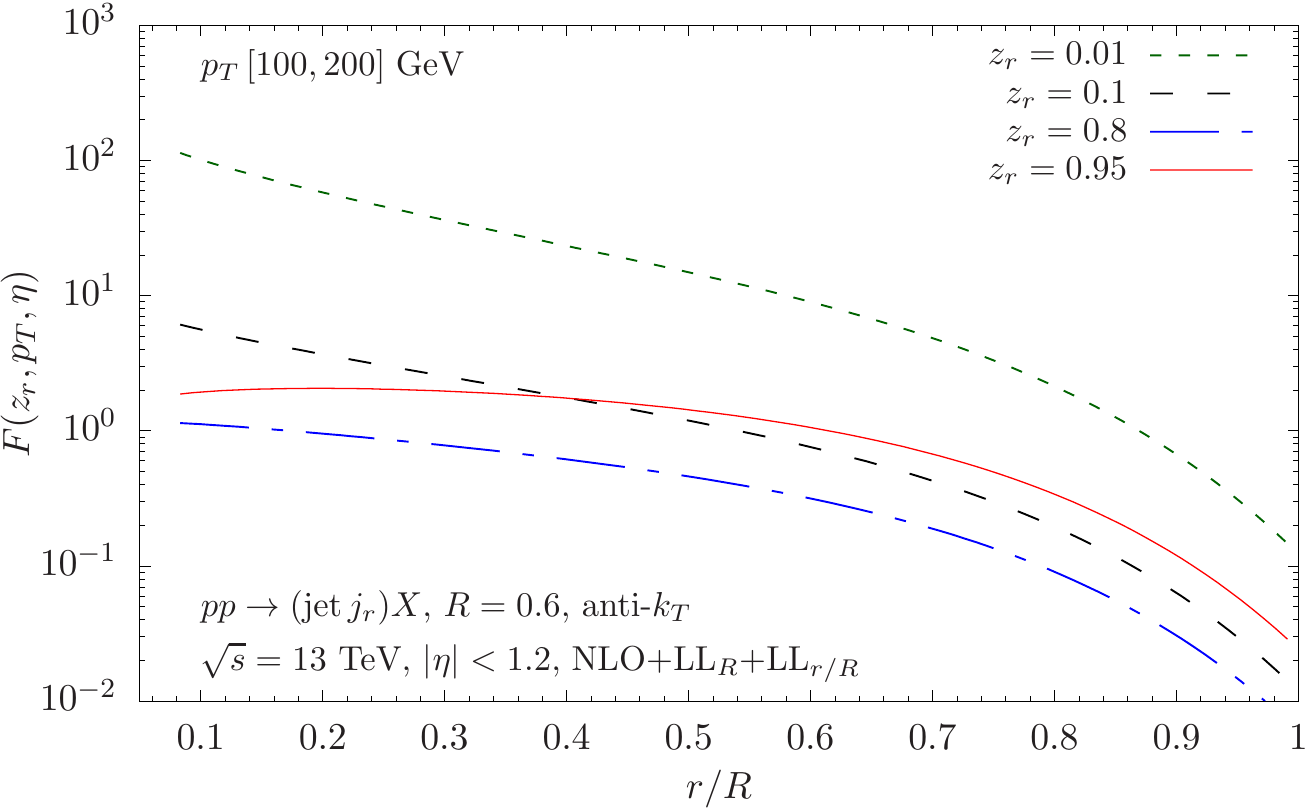} 
\end{center}
\vspace*{1.cm}
\caption{\label{fig:SJF-3} The subjet distribution measured on inclusive jets as a function of $r/R$, using the same kinematics as in \figs{SJF-1}{SJF-2}, and  the jet $p_T$ bin $[25,50]$~GeV (left) and $[100,200]$~GeV (right). Four representative values of the ratio $z_r$ are shown: 0.01 (green dotted), 0.1 (black dashed), 0.8 (blue dot-dashed) and 0.95 (red solid).}
\end{figure*}

Next, we study the dependence of the subjet distribution on the subjet radius parameter $r$. In the right panel of fig.~\ref{fig:SJF-2}, we show the momentum fraction of the subjet for four different values of $r$, ranging from 0.05 to 0.3. We choose the same kinematics as in fig.~\ref{fig:SJF-1} and restrict the jet transverse momentum $p_T$ to the bin $[50,100]$~GeV. We find a relatively strong dependence on $r$ which is also due to the $\ln(r/R)$ resummation effects.  

To make this point more clear, we show in \fig{SJF-3} the dependence of the cross section for fixed values of $z_r$ as a function of $r/R$. Results are shown for four values of $z_r$, and the two panels corresponds to different bins for the jet transverse momenta. One notices again the strong dependence on $r$ which can span two orders of magnitude. For small $z_r$ the curves increase continuously as $r$ decreases, since one finds more and more subjets. However, for sufficiently large $z_r$ the curves flatten out as $r$ becomes small and can even turn over. This behavior is more pronounced for the smaller jet $p_T$ interval of $[25,50]$~GeV, and arises because it is not possible to capture a very large energy fraction $z_r$ of the jet within only a narrow subjet.

\section{Central subjets for the winner-take-all axis} 
\label{sec:central_wta}

In this section we focus on the energy distribution of the subjet centered about the winner-take-all (WTA) axis~\cite{Bertolini:2013iqa}. In \sec{large_r_wta} we treat the case $r \lesssim R$, which parallels the discussion in \sec{subjetfunction}. We discuss the factorization for $r \ll R$ and the resummation of logarithms of $r/R$ in \sec{small_r_wta}. 

\subsection{Central subjet function for $r \lesssim R$}
\label{sec:large_r_wta}

The difference between the standard jet axis and WTA axis resides in the merging step of a clustering algorithm (and is thus not defined for cone jets). Specifically, it chooses the axis to be along the most energetic of the two particles (or pseudojets) that are being merged. For the configuration of at most two partons in the jet, the winner-take-all axis is along the most energetic one. This can simply be accounted for by an additional factor $\theta(z_r - 1/2)$ compared to the $\ord{\al_s}$ calculation in \sec{NLO}. For example, for quark jets
\bea \label{eq:tilde_G_q}
\tilde {\mathcal{G}}^{\mathrm{jet}}_{q}(z,z_r,\omega_R,\mu) 
& = \delta(1-z)\delta(1-z_r) + \f{\as}{2\pi} \delta(1-z_r) L_{R} 
\big[P_{qq}(z)+P_{gq}(z)\big]
\nnu & \quad
+ \f{\as}{2\pi} \delta(1\!-\!z) \theta\Big(z_r-\frac12\Big) L_{r/R}\, [P_{qq}(z_r)+P_{qq}(1\!-\!z_r)] 
\nnu & \quad 
+ \f{\as C_F}{2\pi}
\delta(1\!-\!z_r)
\bigg[\delta(1\!-\!z)  \Big(\f{13}{2}\!-\!\f{2\pi^2}{3}\Big) 
- 2(1+z^2)\left(\f{\ln(1\!-\!z)}{1-z} \right)_+ 
\nn \\ & \quad
- 2\ln(1\!-\!z)\, \frac{1+(1-z)^2}{z} - 1
 \bigg] 
\,, \eea
where the tilde for the SJF indicates that we are restricting to the central subjet about the winner-take-all axis. At higher orders there will be more partons inside the jets, leading to more significant differences between the calculation for the central subjet and an inclusive sample of subjets.

The renormalization of the central subjet function is the same as that of the semi-inclusive jet function. This is immediate at $\ord{\al_s}$ from the above, but holds at higher orders because the rest of the factorization theorem does not depend on whether the energy fractions of the central subjet is measured or not. The central subjet function therefore satisfies the DGLAP evolution equation
\be
\mu \frac{\df}{\df\mu}\, \tilde {\mathcal{G}}_i^{\rm jet}(z,z_r, \omega_R, \mu) = \sum_j \int_z^1\!  \frac{\df z'}{z'}\, \f{\as}{\pi} P_{ji}\Big(\frac{z}{z'} \Big)\, \tilde {\mathcal{G}}_j^{\rm jet}(z',z_r, \omega_R,\mu) \, .
\ee
By evaluating $\tilde {\mathcal{G}}_i$ at its natural scale $\mu_R \sim \omega_R R/2$ and evolving it to  the scale $\mu_H \sim \omega_R/2$ of the hard scattering, the logarithms of $\mu_R/\mu_H \sim R$ are resummed. 

\subsection{Matching for $r \ll R$ and resummation of $\ln(r/R)$}
\label{sec:small_r_wta}

In the regime $r\ll R$, the central subjet function will contain large logarithms of $r/R$ that require resummation. In direct analogy to \eq{matching}, this resummation is accomplished by the following matching equation to all orders in perturbation theory
\be
\label{eq:matching2}
\tilde \GG_i^{\rm jet}(z,z_r,\omega_R,r,R,\mu) = \sum_j \int_{z_r}^1 \frac{\df z_r'}{z_r'} \tilde {\mathcal J}_{ij}(z,z_r',\omega_R,R,\mu)\, \tilde J_j\Big(\frac{z_r}{z_r'},\omega_r,r,\mu\Big) \bigg[1+ \mathcal{O}\bigg(\frac{r^2}{R^2}\bigg)\bigg]\, .
\ee
We first describe this factorization formula and then explain why it holds for the winner-take-all axis.

The object $\tilde J_j$ onto which we match is not the semi-inclusive jet function $J_j$, which describes the distribution of energy fractions for all subjets produced by a parton. Rather, it only picks out the energy fraction of the subjet centered on the winner-take-all axis. It also differs from the central subjet function $\tilde {\mathcal{G}}_i^{\rm jet}$ because all partons are clustered together, since we have effectively taken $R \to \infty$. $\tilde J_j$ is defined as 
\bea
\tilde J_j(z_r, \omega_r, \mu) =& 16\pi^3\,\sum_X \frac{1}{2N_c}\, {\rm Tr} \Big[\frac{\bnslash}{2}
\langle 0| \delta(\omega_R \!-\! \bar n\cdot {\mathcal P}) \delta^2({\mathcal P}_\perp) \chi_n(0)  |X\rangle 
\langle X|\bar \chi_n(0) |0\rangle \Big]
\delta\Big(z_r \!-\! \frac{\omega_r}{\omega_R}\Big)
.\eea
At order $\al_s$ there are at most two partons and the winner-take-all axis is along the most energetic one, so once again
\begin{align}
  \tilde J_j^\one(z_r,\omega_r,\mu) = \theta\Big(z_r > \frac12\Big) J_j^\one(z_r,\omega_r,\mu)
\,.\end{align}
This implies that the one-loop matching coefficients in \eq{matching2} are given by
\begin{align}
\tilde {\mathcal J}_{ij}^\one(z,z_r,\omega_R,\mu) 
&= \Big[\tilde \GG_i^{{\rm jet},\one}(z,z_r,\omega_R,\mu) - \de(1-z)\, \tilde J_j^\one(z_r,\omega_r,\mu)\Big]
\bigg[1+ \mathcal{O}\bigg(\frac{r^2}{R^2}\bigg)\bigg]
\nn \\
&= \theta\Big(z_r > \frac12\Big) \Big[ \GG_i^{{\rm jet},\one}(z,z_r,\omega_R,\mu) - \de(1-z)\,J_j^\one(z_r,\omega_r,\mu) \Big]
\nn \\
& = \theta\Big(z_r > \frac12\Big) {\mathcal J}_{ij}^\one(z,z_r,\omega_R,\mu) 
\,,\end{align}
and thus directly related to those for the inclusive case in \eq{matching}.

In \eq{matching2} the $\tilde \GG_i^{\rm jet}$ on the left-hand side contains physics at angular scales $R$ and $r$, that are factorized into the objects $\tilde {\mathcal J}_{ij}$ and $\tilde J_j$ on the right-hand side. The validity of this equation at next-to-leading order follows immediately from the above. However, to use it for resummation requires the factorization to hold to all orders in $\al_s$. In particular, the axis finding must factorize between the scales $r$ and $R$, i.e.~the axis cannot be sensitive to radiation at the jet boundary. This was shown in ref.~\cite{Neill:2016vbi} for the winner-take-all axis when using Cambridge/Aachen or \kt, in the context of transverse-momentum-dependent fragmentation.

Having performed the factorization in \eq{matching2}, we can now resum the additional logarithms of $r/R$ with the help of another RG equation. The scale $\mu_R \sim \omega_R R/2$ is the natural scale for the matching coefficients $\cJ_{ij}$ and the scale $\mu_r=\omega_r r/2$ is the natural scale for $\tilde J_j$. The evolution of $\tilde J_j$ from $\mu_r$ to $\mu_R$ sums the logarithms of $r/R$, and is described by the following modified DGLAP equation,
\ba \label{eq:tilde_evo}
\mu\f{\df}{\df\mu}\,\tilde J_i^{\rm jet}(z_r,\omega_r,\mu)= \sum_k\int_z^1\f{\df z_r'}{z_r'}\,\tilde \gamma_{ik}\Big(\f{z_r}{z_r'},\mu\Big)\,\tilde J_k^{\rm jet}(z_r',\omega_r,\mu) \, .
\ea
From the NLO expressions in \sec{small_r_wta} it follows that the one-loop anomalous dimensions $\tilde \gamma_{ij}$ are given by
\be
\tilde \gamma_{ij}^\one(z_r, \mu) = \theta\Big(z_r > \frac12\Big)\, \f{\as}{\pi} P_{ji}(z_r) 
\,.\ee

\section{Central subjets for the standard jet axis}
\label{sec:central}

In this section, we discuss the energy distribution of the subjet of radius $r$ centered about the standard jet axis. 
We start with $r \lesssim R$ in \sec{large_r}, which involves a similar calculation as in \sec{NLO}. In \sec{small_r}, we discuss the factorization for  $r \ll R$, which takes on a completely different form than in \secs{matching}{small_r_wta}. In particular, the standard jet axis introduces a sensitivity to (the recoil of) soft radiation. We discuss how the double logarithms of $r/R$ can be resummed in \sec{resum_r}. This factorization suffers from non-global logarithms, obstructing an all-orders resummation.

\subsection{Central subjet function for $r \lesssim R$}
\label{sec:large_r}

We start by introducing the function describing subjets centered about the standard jet axis. To distinguish it from the subjet functions of \secs{subjetfunction}{central_wta} we denote it by $\hat {\mathcal{G}}_i^{\mathrm{jet}}$. We remind the reader that our default notation is for $e^+e^-$ algorithms, and that the central subjet thus corresponds to a cone of opening angle $2r$. On switching to $pp$ algorithms this of course becomes a ``cone" in $(\eta,\phi)$ coordinates.
It is defined by 
\bea
\hat \GG_q^{\mathrm{jet}}(z, z_r, \omega_R, \mu) =& 16\pi^3\,\sum_X \frac{1}{2N_c}\, {\rm Tr} \Big[\frac{\bnslash}{2}
\langle 0| \delta(\omega - \bar n\cdot {\mathcal P}) \delta^2({\mathcal P}_\perp) \chi_n(0)  |X\rangle 
\langle X|\bar \chi_n(0) |0\rangle \Big]
\nnu & \times
\sum_{J_R\in X} \de\Big(z- \frac{\omega_R}{\omega}\Big) \delta\Big(z_r - \frac{\omega_r}{\omega_R}\Big)
\,,\eea
for quark jets, and analogously for gluon jets. There is no sum over subjets $j_r$ in the jet, because we now restrict ourselves to the momentum fraction of the central subjet.

The NLO calculation has the same ingredients as in \sec{NLO}, but the phase-space restrictions for configuration (A) through (C) are modified because we restrict to the subjet centered on the jet axis,
\begin{align} \label{eq:central_nlo}
{\rm (A)} &=
\int\! \df \Phi_2\, \si_{2,i}^c \,
\delta(1-z)\,\delta(1-z_r)\, 
\theta(\beta < R)\, \theta(\beta_1 < r)\, \theta(\beta_2 < r)
\,,\nn \\
{\rm (B)} &=
\delta(1-z)
\int\! \df \Phi_2\, \si_{2,i}^c \,\delta(x - z_r)\, 
\theta(\beta < R)\, \theta(\beta_1 < r)\, \theta(\beta_2 > r)
\,,\nn \\
{\rm (C)} &=
\delta(1-z)
\int\! \df \Phi_2\, \si_{2,i}^c \,\delta(1-x - z_r)\, 
\theta(\beta < R)\, \theta(\beta_1 > r)\, \theta(\beta_2 < r)
\,.\end{align}
The angles $\beta_1, \beta_2$ and $\beta$ were given in \eqs{beta12}{beta}.
The contributions (D) and (E) are not modified. There is also a new (and irrelevant) contribution from the configuration where neither parton is inside the central subjet.
Performing the calculation, and carrying out the renormalization in the $\overline{\rm MS}$ scheme, we find for the cone algorithm
\bea \label{eq:hat_G_cone}
&\hat {\mathcal{G}}^{\text{cone}}_{q}(z,z_r,\omega_R,\mu) 
\nnu
& = \delta(1\!-\!z)\delta(1-z_r)+\f{\as}{2\pi}\bigg\{\delta(1\!-\!z_r)L_R\left[P_{qq}(z)+P_{gq}(z)\right]
+ \delta(1\!-\!z)\theta\bigg(\! z_r \! >\! \f12\!\bigg) L_{r/R} [P_{qq}(z_r)
\nnu & \quad
+P_{gq}(z_r)] +\delta(1\!-\!z)\delta(1\!-\!z_r)C_F\left(\f72\!+\!3\ln 2\!-\!\f{\pi^2}{3} \right)
-\delta(1\!-\!z_r)\bigg[2C_F(1\!+\!z^2)\Big(\f{\ln(1\!-\!z)}{1-z}\Big)_+
\nnu & \quad
 +2P_{gq}(z)\ln(1\!-\!z) +C_F
-2[P_{qq}(z)+P_{gq}(z)]\bigg(\theta\Big(z>\f12\Big) \ln z + \theta\Big(z<\f12 \Big)\ln(1-z)\bigg) \bigg]
\nnu & \quad
-\delta(1-z)\theta\left(\frac{1}{2}\!<\! z_r\! <\! \f{R}{r+R}\right)[P_{qq}(z_r)+P_{gq}(z_r)]\left[L_{r/R}+2\ln\left(\f{1-z_r}{z_r}\right) \right]
\bigg\} ,
 \nnu
&\hat {\mathcal{G}}^{\text{cone}}_{g}(z,z_r,\omega_R,\mu)  
\nnu
&= 
\delta(1\!-\!z)\delta(1\!-\!z_r) \!+\! \f{\as}{2\pi} \bigg\{\delta(1\!-\!z_r)L_{R}
\big[P_{gg}(z)\!+\!2 n_f \,P_{qg}(z)\big]
\!+\! \delta(1\!-\!z)\theta\bigg(\! z_r \! >\! \f12\!\bigg) L_{r/R}\, [P_{gg}(z_r)
 \nnu & \quad 
+2n_f\,P_{qg}(z_r)] 
+ 
\delta(1\!-\!z_r)\delta(1\!-\!z)  \bigg[C_A\Big(\f{137}{36}+\f{11}{3}\ln2\!-\!\f{\pi^2}{3}\Big) -T_F n_f\left(\f{23}{18}+\f43 \ln 2\right)\bigg]
\nnu & \quad
- \delta(1-z_r)\bigg[4C_A\f{(1-z+z^2)^2}{z}\Big(\f{\ln(1\!-\!z)}{1-z} \Big)_+ 
+ 4n_f \big(P_{qg}(z)\ln(1\!-\!z)\, + T_F z(1-z)\big)
\nnu & \quad
-2[P_{gg}(z)+2n_f P_{qg}(z)]\bigg(\theta\Big(z>\f12\Big) \ln z + \theta\Big(z<\f12 \Big)\ln(1-z)\bigg)
 \bigg] 
\nnu & \quad 
 -\delta(1-z)\theta\left(\frac{1}{2}\!<\! z_r\! <\! \f{R}{r+R}\right)[P_{gg}(z_r)+2 n_fP_{qg}(z_r)]\left[L_{r/R}+2\ln\left(\f{1-z_r}{z_r}\right) \right]
\bigg\},
\eea
and for the \kt algorithm
\bea 
&\hat {\mathcal{G}}^{\text{\kt}}_q(z,z_r,\omega_R,\mu) 
\nnu
& = \delta(1\!-\!z)\delta(1\!-\!z_r)+\f{\as}{2\pi}\bigg\{\delta(1\!-\!z_r)L_R\left[P_{qq}(z)+P_{gq}(z)\right]
-\delta(1\!-\!z_r)\bigg[2C_F(1\!+\!z^2)\Big(\f{\ln(1\!-\!z)}{1-z}\Big)_+ 
\nnu & \quad
+2P_{gq}(z)\ln(1-z) +C_F\bigg] 
+ \delta(1-z)\theta\Big(z_r>\f12\Big)[P_{qq}(z_r)+P_{gq}(z_r)](L_{r/R}+2\ln z_r)
\nnu & \quad
+\theta(r<R/2) \bigg[\delta(1-z)\delta(1-z_r)C_F\bigg(\f72+3\ln 2-\f{\pi^2}{3} \bigg)
\nnu & \qquad
-\delta(1-z)\theta\Big(\f12<z_r<1-\f{r}{R} \Big)[P_{qq}(z_r)+P_{gq}(z_r)]\big(L_{r/R}+2\ln(1-z_r)\big)\bigg]
\nnu & \quad
+ \theta(r>R/2) \bigg[
 \delta(1\!-\!z)\delta(1\!-\!z_r)\, C_F\bigg(-\f{1}{2}L_{r/R}^2+\f{3}{2}L_{r/R}-2L_{r/R}\ln\Big(1\!-\!\f{r}{R}\Big)
  +4\text{Li}_2\Big(1\!-\!\f{r}{R}\Big)
\nnu & \qquad 
+\f12-\f{2\pi^2}{3}+6\f{r}{R} \bigg)
-\delta(1-z)\theta\Big(\f12<z_r<\f{r}{R} \Big)[P_{qq}(z_r)+P_{gq}(z_r)]\big(L_{r/R}+2\ln z_r\big)\bigg]
\bigg\} \,,
\nnu
&\hat {\mathcal{G}}^{\text{\kt}}_g(z,z_r,\omega_R,\mu) 
\nnu
& = \delta(1-z)\delta(1-z_r)+\f{\as}{2\pi}\bigg\{\delta(1-z_r)L_R\big[P_{gg}(z)+2n_fP_{qg}(z)\big]
\nnu & \quad
+ \delta(1-z)\theta\Big(z_r>\f12\Big)[P_{gg}(z_r)+2n_fP_{qg}(z_r)](L_{r/R}+2\ln z_r)
\nnu & \quad
-\delta(1-z_r)\bigg[4C_A\f{(1-z+z^2)^2}{z}\Big(\f{\ln(1-z)}{1-z}\Big)_+ +4n_f\big(P_{qg}(z)\ln(1-z) +T_Fz(1-z)\big)\bigg]
\nnu & \quad
 +
 \theta(r<R/2) \bigg[\delta(1-z)\delta(1-z_r)\bigg(C_A\bigg(\f{137}{36}+\f{11}{3}\ln 2-\f{\pi^2}{3}\bigg)-T_F n_f\bigg(\f{23}{18}+\f43\ln 2\bigg) \bigg)
\nnu & \qquad
-\delta(1-z)\theta\Big(\f12<z_r<1-\f{r}{R} \Big)[P_{gg}(z)+2n_fP_{qg}(z)]\big(L_{r/R}+2\ln(1-z_r)\big)\bigg] 
\nnu & \quad
 +\theta(r>R/2) \bigg[\delta(1-z)\delta(1-z_r)\bigg[-\f{C_A}{2}L_{r/R}^2+\f{\beta_0}{2}L_{r/R}-2C_A L_{r/R}\ln\Big(1-\f{r}{R}\Big)
 \nnu & \qquad
 +4 C_A \text{Li}_2\Big(1\!-\!\f{r}{R}\Big)-C_A\f{2\pi^2}{3}+C_A\Big(\f{8r}{R}-\f{r^2}{R^2}+\f{4r^3}{9R^3}\Big)
 + T_F n_f\Big(\f{1}{3}-\f{4r}{R}+\f{2r^2}{R^2}-\f{8r^3}{9R^3}\Big)\!\bigg)
\nnu & \qquad
-\delta(1-z)\theta\left(\f12<z_r<\f{r}{R} \right)[P_{gg}(z)+2n_fP_{qg}(z)]\left(L_{r/R}+2\ln z_r\right)\bigg]\bigg\}\,.
\eea
The renormalization of the central subjet function is again the same as that of the semi-inclusive jet function 
\be
\mu \frac{\df}{\df \mu} \hat {\mathcal{G}}_i^{\rm jet}(z,z_r, \omega_R, \mu) = \sum_j \int_z^1  \frac{\df z'}{z'}\,  \f{\as}{\pi}\, P_{ji}\Big(\frac{z}{z'} \Big)\, \hat {\mathcal{G}}_j^{\rm jet}(z',z_r, \omega_R,\mu) 
\,,\ee
which enables the resummation of logarithms of $R$.

\subsection{Factorization for $r \ll R$}
\label{sec:small_r}

In the regime $r\ll R$, the central subjet function contains large logarithms of $r/R$ that require resummation. This is achieved through a second factorization,
\begin{align} \label{eq:refact}
\hat \GG_i^{\rm jet}(z,z_r,\omega_R,r,R,\mu) &= H_{ij}(z,\omega_R R,\mu)\,
\int\! \df^2 k_\perp\,
 C_j(z_r, \omega_r r, k_\perp,\mu,\nu)\,  S_j(k_\perp,R,\mu,\nu)
 \nn \\ & \quad \times
\Big[1+ \mathcal{O}\Big(\frac{r}{R}, \al_s^2 \ln^2 \frac{r}{R} \Big)\Big]\, ,
\end{align}
where we made the dependence on $r$ and $R$ explicit in the arguments.
We first describe the factorization formula, which differs significantly from \eqs{matching}{matching2}, and then justify it. 

The hard function $H_{ij}$ describes how the energetic parton $i$ produces a jet initiated by parton $j$ with longitudinal momentum $\w_R$ and jet radius $R$, carrying a momentum fraction $z$ of parton $i$. The collinear function $C_j$ describes the fraction $z_r$ of collinear radiation produced by parton $j$, within an angle $r$ of the standard jet axis. It takes into account that the initial collinear parton has a transverse momentum $k_\perp$ with respect to the jet axis, due to the recoil against the soft radiation, encoded in the soft function $S_j$. The transverse momentum dependence causes the factorization in \eq{refact} to suffer from rapidity divergences that require regularization. We will employ the $\eta$-regulator~\cite{Chiu:2011qc,Chiu:2012ir}, for which $\nu$ denotes the corresponding rapidity renormalization scale. Other choices are possible too, see e.g.~refs.~\cite{Collins:1981uk,Dixon:2008gr,Chiu:2009yx,Becher:2011dz,Li:2016axz}.

   \begin{table}
   \centering
   \begin{tabular}{l|c}
     \hline \hline
     Mode: & Scaling $(- , + ,\perp)$  \\ \hline
     hard(-collinear) & $ \w(1,R^2,R)$ \\
     collinear & $ \w(1,r^2,r)$ \\
     (collinear-)soft & $ \w(r/R,r R,r)$ \\ 
     \hline \hline
   \end{tabular}
   \caption{Modes and power counting in SCET that describe the momentum fraction of the central subjet in the regime $r \ll R$.}
   \label{tab:modes}
   \end{table}

The physical justification of \eq{refact} is that the hard(-collinear) radiation cannot undergo a perturbative splitting \emph{inside} the jet. Such a splitting would have a typical opening angle of order $R$ and the contribution of such configurations to the central subjet of radius $r \ll R$ is power suppressed. (Generically, neither of the partons would lie within the central subjets.) Perturbative splittings outside the jet are of course allowed and encoded by the $z$ dependence of the hard function. Collinear splittings inside the jet that affect the central subjet will have typical angle $r$, and are describe by the collinear function. The (collinear-)soft radiation is not energetic enough to influence the $z_r$ measurement, but its transverse momentum $k_\perp$ affects the jet axis, since the total transverse momentum with respect to the jet axis is zero, and must be taken into account. In the language of SCET, these correspond to distinct degrees of freedom with the parametric scaling of momenta summarized in table~\ref{tab:modes}.

The collinear function has the following definition for $j=q$,
\bea
 C_q(z_r, \omega_r r, k_\perp,\mu,\nu) =& 16\pi^3\,\sum_X \frac{1}{2N_c}\, {\rm Tr} \Big[\frac{\bnslash}{2}
\langle 0| \delta(\omega_R \!-\! \bar n\cdot {\mathcal P}) \delta^2({\mathcal P}_\perp \!-\! k_\perp) \chi_n(0)  |X\rangle 
\langle X|\bar \chi_n(0) |0\rangle \Big]
\nnu & \times
\delta\Big(z_r - \frac{\omega_r}{\omega_R}\Big)
\,,\eea
and similarly for $j=g$.
This describes the momentum fraction $z_r$ of the central subjet centered on the $n$ axis. The recoil of the collinear radiation with respect to the jet axis due to soft radiation is taken into account through the $\delta^2({\mathcal P}_\perp \!-\! k_\perp)$. 

The definition of the soft function for $j=q$ is given by
\begin{align} \label{eq:S_def}
S_q(k_\perp, R, \mu, \nu)
=  \frac{1}{N_c} \sum_{X_s} \langle 0  |  {\rm\bar T}[Y_{\bn}^\dagger Y_n]\, |X\rangle \langle X | {\rm T}[Y_n^\dagger Y_{\bn} ]  |  0  \rangle \de\bigg(k_\perp - \sum_{i\in X} \theta(\beta_i < R)\, k_{i,\perp}\bigg)
\,.\end{align}
The delta function sums the transverse momentum $k_{i,\perp}$ of soft radiation inside the jet, $\beta_i <R$. $Y_n$ is a soft Wilson line in the fundamental representation along the light-like direction $n^\mu = (1, \hat n)$ of the jet, 
\begin{align}
  Y_n(x) = {\rm \bar P} \exp\Big[-\img g \int_0^\infty\! \df t\, n \sdt A_s(t\, n^\mu) \Big]
\end{align}
and $Y_{\bn}$ is along the opposite direction $\bn^\mu = (1, - \hat n)$.  For $j=g$ the Wilson lines are in the adjoint representation and the overall normalization is modified $1/N_c \to 1/(N_c^2-1)$. The hard function $H_{ij}$ does not have a direct matrix element definition in SCET, but instead it is defined by the matching relation in \eq{refact}.

The factorization for the standard jet axis in \eq{refact} does not account for non-global logarithms (NGLs)~\cite{Dasgupta:2001sh,Dasgupta:2002dc}. These arise because the transverse momentum of the (collinear-)soft radiation inside the jet is probed, but is unconstrained outside the jet.\footnote{Boosting to the frame where the jet becomes a hemisphere, the modes in table~\ref{tab:modes} become the standard SCET$_{\rm II}$ hard, collinear and soft modes. Emissions into the other hemisphere are unconstrained and lead to additional (collinear-)soft Wilson lines~\cite{Larkoski:2015zka,Becher:2016mmh}. In the original frame these corresponds to emissions outside the jet described by $H_{ij}$.} 

The tree-level hard, collinear and soft functions are given by
\begin{align}
  H_{ij}(z,\w_R R,\mu) &= \de_{ij} \de(1-z)
  \,, \nn \\ 
  C_j(z_r,\w_r r,k_\perp, \mu,\nu) &= \de(1-z_r) \theta\Big(|k_\perp| < \frac{\w_r r}{2}\Big)
  \,, \nn \\
  S_j(k_\perp,R, \mu,\nu) & = \de^2(k_\perp)
\,.\end{align}

We calculate the one-loop corrections in pure dimensional regularization, such that the virtual corrections are scaleless and vanish. The contributions to $H_{ij}$ come from perturbative splittings of the parton $i$ where the jet consists solely of parton $j$. For $H_{qq}^\one$ and $H_{qg}^\one$ these can directly be read off from diagrams (D) and (E) in \sec{NLO}. For the \kt algorithm, we find
\begin{align}
 H_{qq}^{\one,\text{\kt}}(z,\w_R R,\mu) 
  &= \f{\as}{2\pi} \bigg[C_F \delta(1-z)\Big(-\f{L_{R}^2}{2} - \frac32 L_R +\f{\pi^2}{12} \Big) 
\nnu
 & \quad
+L_{R} P_{qq}(z) -2C_F(1+z^2)\Big(\f{\ln(1-z)}{1-z}\Big)_+ -C_F(1-z)  \bigg] 
, \nnu
 H_{qg}^{\one,\text{\kt}}(z,\w_R R,\mu) 
 &=\f{\as}{2\pi}\bigg[\Big(L_{R} - 2 \ln(1-z) \Big) P_{gq}(z) - C_Fz \bigg]
, \nnu
H_{gq}^{\one,\text{\kt}}(z,\w_R R, \mu) 
 & =  \f{\as}{2\pi}\bigg[\Big(L_{R} - 2\ln(1-z) \Big)  P_{qg}(z) - T_F 2z(1-z) \bigg]
, \nnu
H_{gg}^{\one,\text{\kt}}(z, \w_R R, \mu) 
& = \f{\as}{2\pi}\bigg[ \delta(1-z)\Big(-C_A\f{L_R^2}{2} - \f{\beta_0}{2} L_R + C_A \frac{\pi^2}{12}\Big)
\nnu 
& \quad
+ L_R P_{gg}(z) - \frac{4C_A (1-z+z^2)^2}{z} \left(\frac{\ln(1-z)}{1-z}\right)_{+} \bigg]
.\end{align}
Similarly, the results for the cone algorithm can be written as
\begin{align} \label{eq:H_cone}
H_{ij}^{\one,\text{cone}}(z,\w_R R,\mu)=&H_{ij}^{\one,\text{\kt}}(z,\w_R R,\mu)
\nnu
&
+\f{\as}{2\pi} 2P_{ji}(z)\left[\theta\Big(z>\f12 \Big)\ln z+\theta\Big(z<\f12\Big)\ln(1-z)\right].
\end{align}

We next consider the soft function, which measures the transverse momentum of soft radiation in the jet. Performing the calculation using ref.~\cite{Kasemets:2015uus}, and noting that the jet region corresponds to rapidity $y>-\ln (R/2)$ with respect to the \emph{jet axis}, we obtain
\begin{align} \label{eq:S_nlo}
  S_q^\one(k_\perp,R, \mu,\nu) =  \frac{\al_s C_F}{2\pi^2} \bigg[-  \frac{1}{\mu^2}\,
\Big( \frac{\ln(k_\perp^2/\mu^2)}{k_\perp^2/\mu^2}\Big)_+\!
+ \frac{1}{\mu^2}\,
 \frac{1}{(k_\perp^2/\mu^2)}_+\!\ln \frac{\nu^2 R^2}{4\mu^2}
 - \frac{\pi^2}{12} \de(\vec k_\perp^{\,2})\bigg]
\,.\end{align}
The result for $S_g^{(1)}$ follows by replacing $C_F \to C_A$.

The full collinear function is already complicated at NLO, because it involves two measurements.\footnote{This seems similar to the case of jet broadening, for which the collinear contribution at one loop was only calculated in ref.~\cite{Becher:2012qc}.} As our current approach is anyway limited to NLL order due to non-global logarithms, we simply consider $k_\perp = 0$ (from the tree-level soft function). The calculation of the collinear function involves a slight modification to contributions (A), (B) and (C) to the central subjet function in \sec{large_r}. Since $r \ll R$, the collinear radiation is close to the center of the jet and does not probe the jet boundary, removing the $\theta(\beta <R)$ and $\de(1-z)$ in \eq{central_nlo}. For the quark case, the individual contributions are given by
\begin{align}
{\rm (A)} &=
\int\! \df \Phi_2\, \si_{2,i}^c \,
\delta(1-z_r)\, 
 \theta(\beta_1 < r)\, \theta(\beta_2 < r)
 \nn \\
 &= \frac{\al_s C_F}{2\pi}\, \de(1-z_r) \Big[\frac{1}{\eps^2} + \frac{3}{2\eps} + \frac{L_r}{\eps} + \frac{L_r^2}{2}+\frac{3}{2}L_r+\frac{7}{2}-\frac{5\pi^2}{12}+3\ln 2 +\ord{\eps} \Big]
\,,\nn \\
{\rm (B)} &=
\int\! \df \Phi_2\, \si_{2,i}^c \,\delta(x - z_r)\, 
\theta(\beta_1 < r)\, \theta(\beta_2 > r)\, \Big(\frac{\nu}{(1-x)\w_R}\Big)^\eta
\nn \\
&= \frac{\al_s}{2\pi}\, 
\bigg[\de(1-z_r)C_F\bigg( \frac{2}{\eta}\Big(\frac{1}{\eps}+L_r\Big) - \frac{1}{\eps^2}
+\frac{1}{\eps}(2  \ln \frac{\nu}{\w_R} - L_r) - \frac{L_r^2}{2} + 2  \ln \frac{\nu}{\w_R} L_r  + \frac{\pi^2}{12}\bigg)
\nn \\ & \quad +
\theta\Big(z_r > \frac12\Big) \bigg(
-4C_F\Big(\frac{\ln(1\!-\!z_r)}{1-z_r}\Big)_+ \!\!+\!2C_F(1\!+\!z_r) \ln(1\!-\!z_r) \!+\! 2P_{qq}(z_r) \ln z_r\bigg) \!+\! \ord{\eta,\eps}
\bigg]
,\nn \\
{\rm (C)} &=
\int\! \df \Phi_2\, \si_{2,i}^c \,\delta(1-x - z_r)\, 
\theta(\beta_1 > r)\, \theta(\beta_2 < r)
\nn \\
&= \frac{\al_s}{2\pi}\, \theta\Big(z_r > \frac12\Big)
\Big[2P_{gq}(z_r) \ln \Big(\frac{z_r}{1-z_r}\Big) + \ord{\eps}\Big]
\,.\end{align}
Here we needed to include the $\eta$-regulator for contribution (B).
Adding up the various contributions and performing the renormalization, 
\begin{align}
C_q(z_r, \omega_r r, 0,\mu,\nu) 
 &= {\rm (A)} + {\rm (B)} + {\rm (C)}
\nn \\
&= \frac{\al_s}{2\pi}\, 
\bigg\{\de(1-z_r)C_F\bigg[
  \Big(2  \ln \frac{\nu}{\w_R} + \frac32\Big) L_r  + \frac{7}{2}-\frac{\pi^2}{3}+3\ln 2  \bigg]
\nn \\ & \quad +
\theta\Big(z_r > \frac12\Big) \bigg[
-2C_F(1+z_r^2)\Big(\frac{\ln(1-z_r)}{1-z_r}\Big)_+ -2P_{gq}(z_r) \ln (1-z_r)
\nn \\ & \quad
 + 2\big(P_{qq}(z_r) +P_{gq}(z_r) \big) \ln z_r\bigg] \bigg\}
.\end{align}
A similar calculation yields the gluon collinear function
\begin{align}
C_g(z_r, \omega_r r, 0,\mu,\nu) &= \frac{\al_s}{2\pi}\, 
\bigg\{\de(1 \!-\! z_r)\bigg[
  \Big(2 C_A \ln \frac{\nu}{\w_R} \!+\! \frac{\beta_0}{2}\Big) L_r  \!+\! \Big(\frac{7}{24} \!-\! \frac{\pi^2}{3}\Big)C_A \!+\! \Big(\frac{23}{24} \!+\!  \ln 2\Big)\beta_0  \bigg]
\nn \\ & \quad +
\theta\Big(z_r > \frac12\Big) \bigg[
-4C_A\f{(1-z_r+z_r^2)^2}{z_r}\Big(\frac{\ln(1\!-\!z_r)}{1\!-\!z_r}\Big)_+ \!\!-2P_{qg}(z_r) \ln (1\!-\!z_r)
\nn \\ & \quad
+ 2\big(P_{gg}(z_r) +P_{qg}(z_r)\big) \ln z_r\bigg]  \bigg\}
.\end{align}

We have verified that these one-loop ingredients indeed satisfy \eq{refact}. This check involved a subtlety related to distributions: Since
\begin{align}
\theta\Big(\frac{1}{2}<z_r<\f{R}{r+R}\Big) = 
\theta\Big(\frac{1}{2}<z_r<1\Big) \Big[1 + \mathcal{O}\Big(\frac{r}{R}\Big)\Big]
\,,\end{align}
a naive expansion of $\hat \cG_i^{\rm cone}$ in $r/R$ leads to improperly regulated plus distributions such as $\ln(1-z_r) /(1-z_r)_+$ instead of $[\ln(1-z_r)/(1-z_r)]_+$. Rather, we verify \eq{refact} by first considering $z_r<1$ and then taking a suitable integral containing the point $z_r=1$,  \emph{before} expanding in $r/R$. Note that the difference between $\hat \cG_i^{\rm cone}$ and $\hat \cG_i^\text{\kt}$ is completely captured by \eq{H_cone} for $r \ll R$.

\subsection{Resummation of $\ln (r/R)$}
\label{sec:resum_r}

The resummation is achieved by evaluating each of the ingredients in \eq{refact} at their natural scale
\begin{align}
  \mu_H &\sim p_T R\,,
  \nn \\
  \mu_C &\sim p_T\, r\,, \quad
  \nu_C \sim p_T\,,
  \nn \\
  \mu_S &\sim p_T\, r\,, \quad
  \nu_S \sim p_T\, \frac{r}{R}\,,
\end{align}
and evolving them to common scales $\mu$ and $\nu$ using their RG equations
\begin{align}
\mu\f{\df}{\df\mu}\,H_{ij}(z,\omega_R R,\mu) &= \sum_k \int_z^1\f{\df z'}{z'}\,\gamma^H_{ik}\Big(\f{z}{z'},\w_R R,\mu\Big)\,H_{kj}(z',\omega_R R,\mu)
\,. \nn \\
\mu\f{\df}{\df\mu}\,C_i(z_r,\omega_r r,k_\perp,\mu,\nu) &= \gamma^C_i(\w_R,\mu, \nu)\,C_i(z_r,\omega_r r,k_\perp,\mu,\nu)
\,. \nn \\
\mu\f{\df}{\df\mu}\,S_i(k_\perp,R, \mu,\nu) &= \gamma^S_i(\mu, \nu R)\,S_i(k_\perp,R, \mu,\nu)
\,, \nn \\
\nu\f{\df}{\df\nu}\,C_i(z_r,\omega_r r,k_\perp,\mu,\nu) &= -\gamma^\nu_i(k_\perp,\nu)\,C_i(z_r,\omega_r r,k_\perp,\mu,\nu)
\,. \nn \\
\nu\f{\df}{\df\nu}\,S_i(k_\perp,R, \mu,\nu) &= \gamma^\nu_i(k_\perp,\mu)\,S_i(k_\perp,R, \mu,\nu)
\,. \end{align}
To avoid a cumbersome complication for resummation in momentum space~\cite{Frixione:1998dw}, it is much more convenient to carry out the rapidity resummation in impact parameter space.\footnote{Recent work has shown how to carry out this resummation directly in momentum space~\cite{Monni:2016ktx,Ebert:2016gcn}.}

The one-loop anomalous dimensions directly follow from the expressions in \sec{small_r},
\begin{align}
 \ga^H_{qq}(z,\w_R R,\mu)
  &= \f{\as}{\pi} \bigg[C_F \Big(-  L_R - \frac32\Big)\, \delta(1-z)
+ P_{qq}(z)  \bigg]
, \nn \\
 \ga^H_{qg}(z,\w_R R,\mu) 
 &= \f{\as}{\pi}\, P_{gq}(z)
, \nn \\
\ga^H_{gg}(z,\w_R R,\mu)
  &= \f{\as}{\pi} \bigg[\Big(- C_A  L_R - \frac12 \beta_0 \Big)\, \delta(1-z)
+ P_{gg}(z)  \bigg]
, \nn \\
 \ga^H_{gq}(z,\w_R R,\mu) 
 &= \f{\as}{\pi}\, P_{qg}(z)
, \nn \\
\ga^C_q(\omega_R,\mu,\nu) 
&= \frac{\al_s C_F}{\pi}\, 
  \Big(2  \ln \frac{\nu}{\w_R} + \frac32\Big)
\,, \nn \\
\ga^C_g(\omega_R,\mu,\nu) 
&= \frac{\al_s}{\pi}\, 
  \Big(2C_A  \ln \frac{\nu}{\w_R} + \frac12 \beta_0\Big)
\,, \nn \\
\ga^S_q(\mu,\nu R)
&= \frac{\al_s C_F}{\pi}\, \ln \frac{4\mu^2}{\nu^2 R^2}
\,, \nn \\
\ga^S_g(\mu,\nu R)
&= \frac{\al_s C_A}{\pi}\, \ln \frac{4\mu^2}{\nu^2 R^2}
\,, \nn \\
\ga^\nu_q(k_\perp,\mu)
&= \frac{\al_s C_F}{\pi}\, \frac{1}{\mu^2}\,\frac{1}{(k_\perp^2/\mu^2)}_+
\,, \nn \\
\ga^\nu_g(k_\perp,\mu)
&= \frac{\al_s C_A}{\pi}\, \frac{1}{\mu^2}\,\frac{1}{(k_\perp^2/\mu^2)}_+
\,.\end{align}
The anomalous dimensions of the hard, collinear and soft function do not combine to zero, as they should yield the anomalous dimension of the central subjet function, see \eq{refact}. This is indeed the case, since the splitting function contributions in $\ga^H_{ij}$ remain uncancelled.

\section{The jet shape}
\label{sec:jetshape}

In this section, we consider the jet shape which is the average momentum fraction of the central subjet distribution. For comparison, we also consider the inclusive subjet sample.

\subsection{Inclusive subjets}

The average momentum fraction for inclusive subjets simply amounts to a sum rule, in contrast to the case of central subjets. Specifically, averaging the SJFs ${\cal G}_i^{\rm jet}$ over $z_r$, we get back the semi-inclusive jet functions
\bea
\int_0^1\df z_r\,z_r\, {\cal G}_{i}^{\rm jet}(z, z_r, \omega_R, \mu) = J_i(z, \omega_R, \mu)
\,. \eea
This result holds both for all jet algorithms and for $i=q,g$, as it is simply due to momentum conservation. 

Applying this to the $r \ll R$ limit, described by the matching in \eq{matching}, we find the following
\bea \label{eq:mom_matching}
\int_0^1\df z_r\,z_r\, {\cal G}_{i}^{\rm jet}(z, z_r, \omega_R, \mu) = 
\sum_j \int_0^1\df z_r\,z_r\, {\cal J}_{ij}(z,z_r,\omega_R,\mu) 
\int_0^1\df z'\,z' J_j(z',\omega_r,\mu) 
\,. \eea
This equation can be verified by using the momentum sum rule for the siJF in \eq{sumrule} and combining it with the momentum sum rule for the fragmenting jet function~\cite{Jain:2011xz,Kang:2016ehg} 
\bea \label{eq:sumrule2}
\int_0^1\df z_r\,z_r\big[{\cal J}_{qq}(z,z_r,\omega_R,\mu)+{\cal J}_{qg}(z,z_r,\omega_R,\mu)\big] &= J_q(z,\omega_R,\mu) \, ,
\nn \\
\int_0^1\df z_r\,z_r\big[{\cal J}_{gg}(z,z_r,\omega_R,\mu)+2n_f{\cal J}_{gq}(z,z_r,\omega_R,\mu)\big] &= J_g(z,\omega_R,\mu) \,.
\eea
In particular, by averaging over $z_r$, all logarithms $L_{r/R}$ in ${\cal G}_{i}^{\rm jet}$ disappear. This will not be the case for central subjets, as discussed in the following sections.

\subsection{Winner-take-all axis}

The $z_r$ averaged results for the SJFs for central subjets along the WTA axis are given by:
\bea
\int_0^1\df z_r\,z_r\, \tilde {\cal G}_{q}^{\rm jet}(z, z_r, \omega_R, \mu) & = J_q(z, \omega_R, \mu) +\delta(1-z)\,\f{\as C_F}{2\pi}\, L_{r/R}\left(\f{3}{8}-2\ln 2\right)
\,,\\
\int_0^1\df z_r\,z_r\, \tilde {\cal G}_{g}^{\rm jet}(z, z_r, \omega_R, \mu) & = J_g(z, \omega_R, \mu) +\delta(1-z)\,\f{\as}{2\pi}\,L_{r/R}\bigg[C_A\Big(\f{43}{96}-2\ln 2\Big)-T_F n_f \f{7}{48}\bigg]
.\nn \eea
In this case we have a single logarithmic dependence on $L_{r/R}$. The above expressions hold for both the cone and the anti-k$_T$ algorithm, as the algorithm-dependent pieces of $\tilde{\cal G}_i^{\rm jet}$ are contained in $J_i$.

For $r \ll R$, we have
\bea 
\int_0^1\df z_r\,z_r \tilde {\cal G}_{i}^{\rm jet}(z, z_r, \omega_R, \mu) = 
\sum_j \int_0^1\df z_r\,z_r \tilde {\cal J}_{ij}(z,z_r,\omega_R,\mu) 
\int_0^1\df z'\,z' \tilde J_j(z',\omega_r,\mu) 
\,. \eea
in direct analogy with \eq{mom_matching}. However, the sum rules in \eqs{sumrule}{sumrule2} do not hold for  $\tilde J_j$ and $\tilde {\cal J}_{ij}$ due to the additional theta functions, which is why the single logarithms of $r/R$ persist.

\subsection{Standard jet axis}

Here we present the $z_r$ averaged results for subjets along the standard jet axis, showing separate results for the cone and anti-k$_T$ algorithm. We obtain
\bea\label{eq:central_zraverage}
\int_0^1\df z_r\,z_r\, \hat {\cal G}_{q}^{\text{\kt}}(z, z_r, \omega_R, \mu) & = J^{\text{\kt}}_q(z, \omega_R, \mu) + \delta(1-z)\,\f{\as C_F}{2\pi} \bigg[-\f{1}{2}L_{r/R}^2+\f{3}{2}L_{r/R}-\f{9}{2}
\nn\\ &\quad 
+\frac{6r}{R}-\f{3r^2}{2R^2} \bigg]
\,, \nnu
\int_0^1\df z_r\,z_r\, \hat {\cal G}_{g}^{\text{\kt}}(z, z_r, \omega_R, \mu) & = J^{\text{\kt}}_g(z, \omega_R, \mu) + \delta(1-z) \f{\as}{2\pi} \bigg[ - \frac{C_A}{2}L_{r/R}^2 + \f{\beta_0}{2} L_{r/R} 
\nn \\ & \quad
+C_A\Big( -\f{203}{36} +\f{8r}{R} - \f{3r^2}{R^2} + \f{8r^3}{9R^3} - \f{r^4}{4R^4}\Big)
\nn \\ & \quad
+ T_F n_f\Big(\f{41}{18} - \f{4r}{R} + \f{3r^2}{R^2} - \f{16r^3}{9R^3} + \f{r^4}{2R^4}\Big)\bigg]
\,, \nnu
\int_0^1\df z_r\,z_r\, \hat {\cal G}_{q}^{\rm cone}(z, z_r, \omega_R, \mu) & = J^{\rm cone}_q(z, \omega_R, \mu) + \delta(1-z)\f{\as C_F}{2\pi} \bigg[-\f{1}{2}L_{r/R}^2+\f{3}{2}L_{r/R}+\f{3}{2}
\nn\\ & \quad
 -3\ln 2-\f{\pi^2}{3}+4{\rm Li}_2\left(\f{r}{r+R}\right)+2\ln^2\Big(1+\frac{r}{R}\Big)
\nn\\ & \quad
+3\ln\Big(1+\frac{r}{R}\Big)+\f{3r}{r+R}\bigg]
\,, \nnu
\int_0^1\df z_r\,z_r\, \hat {\cal G}_{g}^{\rm cone}(z, z_r, \omega_R, \mu) & = J^{\rm cone}_g(z, \omega_R, \mu) + \delta(1-z)\f{\as}{2\pi} \left[-\f{C_A}{2}L_{r/R}^2+\f{\beta_0}{2}L_{r/R}\right.
\nn\\ & \quad
+2C_A\ln^2\left(1+\f{r}{R}\right)+\beta_0\ln\left(1+\f{r}{R}\right) +4C_A \,{\rm Li}_2\left(\f{r}{r+R}\right)
\nn\\ & \quad
-\beta_0\ln 2-C_A\f{\pi^2}{3}+ C_A\f{R-r}{6(r+R)^3}\left(11r^2+22 rR+12R^2\right)
\nn\\ & \quad
\left. + \,T_F n_f \f{R-r}{3(r+R)^3} \left(2r^2+4rR+3R^2\right) \right]
\,. \eea
Note that here we have a double logarithms of $L_{r/R}$ and that these expressions contain power corrections of the form $r/R$. We have compared this with results available in the literature: Combining the in-jet calculation of ref.~\cite{Chien:2014nsa} (see also refs.~\cite{Seymour:1997kj,Li:2011hy,Li:2012bw} for earlier results obtained within standard QCD) with the out-of-jet contribution of ref.~\cite{Kang:2016mcy}, we find agreement with \eq{central_zraverage}. Also, for $r=R$ these results reduce to the semi-inclusive jet function in ref.~\cite{Kang:2016mcy}.

Our refactorized cross section in the limit $r\ll R$ and, hence, the resummation of logarithms $L_{r/R}$ for the jet shape takes on a different form than in the literature. From \eq{refact} it follows that
\begin{align}
\int_0^1\df z_r\,z_r\, 
\hat \GG_i^{\rm jet}(z,z_r,\omega_R,r,R,\mu) &= H_{ij}(z,\omega_R,R,\mu)\,
\int\! \df^2 k_\perp\,
 \Big[\int_0^1\df z_r\,z_r\, C_j(z_r, \omega_r r, k_\perp,\mu,\nu)\Big]\,  
 \nn \\ & \quad \times S_j(k_\perp,R, \mu,\nu)
\Big[1+ \mathcal{O}\Big(\frac{r}{R}, \al_s^2 \ln^2 \frac{r}{R} \Big)\Big]\,,
\end{align}
where the refactorization of the soft (recoil) contribution is the new ingredient. This additional factorization is essential to resum the logarithms of $r/R$ beyond LL accuracy. 

\subsection{Relation with TMD fragmentation}

Because averaging is linear, the jet shape can directly be related to TMD fragmentation through the following sum rule 
\begin{align} \label{eq:TMDfrag}
   \int\! \df z_r\, z_r\, \tilde {\mathcal{G}}^{\mathrm{jet}}_{i}(z,z_r,\omega_R,\mu)
    = 
    \sum_h \int_{|k_\perp| \leq \w_R r/2} \df^2 k_\perp \,
     \int\! \df z_h\, z_h\, \tilde {\mathcal{G}}^h_{i}(z,\omega_R,k_\perp, z_h, \mu)
\,,\end{align}
and similarly for the standard jet axis (replacing tildes by hats).
This formula describe the central subjet as the sum of the contributions of its hadron constituents. The TMD fragmentation function $\tilde {\mathcal{G}}^h_{i}(z,\omega_R, k_\perp, z_h, \mu)$ is the number density of hadrons of species $h$, momentum fraction $z_h$ and transverse momentum $z_h k_\perp$ with respect to the winner-take-all axis~\cite{Neill:2016vbi}. The restriction to the central subjet of radius $r$ is encoded by 
\be
  r \geq \theta_h \approx \frac{z_h |k_\perp|}{z_h \w_R/2} = \frac{2|k_\perp|}{\w_R} 
\,.\ee
We have verified that \eq{TMDfrag} holds for \kt with the winner-take-all axis as well as the standard jet axis using the one-loop results in refs.~\cite{Neill:2016vbi} and \cite{Bain:2016rrv,Kang:2017glf}.

\section{Conclusions}
\label{sec:conclusions}

In this paper we considered the energy fraction $z_r$ of subjets of size $r$ inside a jet of size $R$. We presented analytical results for the following three cases: inclusive subjets obtained by reclustering all particles in the jet with jet radius parameter $r$, as well as central subjets along the winner-take-all axis and along the standard jet axis. The single logarithms of the form $\as^n\ln^n R$ are the same in each case and can be resummed to all orders by solving the associated DGLAP evolution equation. 

We also considered the logarithms of the ratio of the jet size parameters $\ln(r/R)$, whose structure depends on the particular subjet observable. For each case, we performed an additional refactorization of the cross section in the limit $r\ll R$, enabling the resummation for this class of logarithms. For central subjets along the WTA axis, this refactorization is known to all-orders in $\al_s$ but we are currently restricted to leading logarithmic resummation because of our knowledge of the anomalous dimensions. For central subjets along the standard jet axis, an all-orders factorization formula is hindered by non-global logarithms. We presented numerical results for the $z_r$-distribution of inclusive subjets measured on an inclusive jet sample $pp\to(\mathrm{jet}\, j_r)X$, leaving numerical results for central subjets to future work~\cite{futurejetshapes}. 
In addition, we considered the average energy fraction of these results, which is known as the jet shape for subjets centered on the standard jet axis. For the jet shape, our factorization formula in the limit $r\ll R$ involves an additional refactorization compared to the literature, to account for the recoil effect of soft radiation on the jet axis. Along the way, we also pointed out an inconsistency in the literature for analytical results for inclusive cone jets and their substructure. 

There are various possible applications of our work in the future. First of all, it will be very interesting to perform numerical calculations for central subjets along the two axes we considered in this work. This will be particularly relevant since our factorization for the standard jet shape in the limit $r\ll R$ differs from that in the available literature at next-to-leading logarithmic order, and it is possible to compare to experimental data in this case. In addition, for the more differential case (the $z_r$-dependent case), experimental measurements will be feasible which can shed new light on the substructure of jets at the LHC. In addition, it will be interesting to further explore the possibility of how these ``relatively inclusive'' jet substructure observables can be used to discriminate between QCD jets and boosted objects. Finally, we expect that the different energy distributions of subjets considered in this work can be very relevant to better understand the properties of the quark-gluon plasma created in heavy-ion collisions.

\acknowledgments
We thank Y. -T.~Chien, H.~n.~Li, X.~Liu, D.~Neill, V.~Rentala and G.~Sterman for discussions. We also thank  T. Kaufmann, I. Vitev, and W. Vogelsang for comments on the manuscript. This work is supported by the National Science Foundation under Contract No. PHY-1720486, the U.S. Department of Energy under Contract Nos.~DE-AC02-05CH11231, DE-AC52-06NA25396, by the Laboratory Directed Research and Development Program of Lawrence Berkeley National Laboratory, by the ERC grant ERC-STG-2015-677323 and the D-ITP consortium, a program of the Netherlands Organization for Scientific Research (NWO) that is funded by the Dutch Ministry of Education, Culture and Science (OCW).

\bibliographystyle{JHEP}
\bibliography{bibliography}

\end{document}